\documentclass[final,1p,times]{elsarticle}

\usepackage{amssymb, amsmath}

\usepackage{graphicx,epsf}
\usepackage{color}

\journal{Nuclear Physics B}

\begin{document}

\begin{frontmatter}


\title{Power-law out of time order correlation functions in the SYK model} 

\author[label1]{Dmitry Bagrets}
\author[label1]{Alexander Altland}
\author[label2]{Alex Kamenev}
\address[label1]{Institut f\"ur Theoretische Physik, Universit\"at zu K\"oln,
Z\"ulpicher Stra\ss e 77, 50937 K\"oln, Germany}
\address[label2]{W. I. Fine Theoretical Physics Institute and School of Physics and Astronomy, University
of Minnesota, Minneapolis, MN 55455, USA}


\begin{abstract}
We  evaluate the finite temperature partition sum and correlation functions of the 
Sachdev-Ye-Kitaev (SYK) model. Starting from a recently proposed mapping of the SYK model  onto  
Liouville quantum mechanics, we obtain our results by  exact integration over conformal Goldstone modes  
reparameterizing physical time.  Perhaps, the least expected result of our analysis is that at time scales proportional to the number of particles the  out of time order correlation function crosses over from a regime of  exponential decay to a universal $t^{-6}$ power-law behavior. 
\end{abstract}


\end{frontmatter}


\section{Introduction}
\label{sec:Intro}

The Sachdev-Ye-Kitaev (SYK) model \cite{Sachdev:1993,Kitaev:2015} is a system of $N$ (Majorana) fermions, $\chi_i$ with $i=1,\ldots,N$, subject to a four-fermion interaction
\begin{align}
    \label{eq:Hamiltonian}
    \hat H=\sum_{ijkl}^N J_{ijkl}\,\chi_i\chi_j\chi_k\chi_l,
\end{align}
with Gaussian distributed random matrix elements $J_{ijkl}$ with zero mean and a variance given by  $\langle |J_{ijkl}|^2\rangle=6J^2/N^{3}$. The seeming simplicity of this model
is deceptive. At low excitation energies it exhibits an asymptotically exact
conformal symmetry \cite{Kitaev:2015,Sachdev:2015,Maldacena:2016,Polchinski:2016}  which manifests itself in the infinite-dimensional freedom to
re-parameterize time, $t\to f(t)$, in the description of long time correlations. This
`nearly conformal symmetry' (NCFT)~\cite{Maldacena:2016,Maldacena1:2016} makes the model a candidate
holographic shadow of some two-dimensional bulk. The potential realization of a
holographic principle of lowest possible dimension has triggered a surge of research
activity on the SYK model and its quantum dynamics~\cite{Banerjee:2016, Berkooz:2016, Davidson:2016, Cotler:2016, Danshita:2016, Garcia:2016, Garcia:2017, Engelsoy:2016, Fu:2017, Jensen:2016, Pikulin:2017, Turiaci:2017, Witten:2016, You:2016}.

At the same time, the existence of a large $N$ parameter within an `infinite range'
interaction framework make the system amenable to mean-field approaches. It turns out
that at the mean-field level  the infinite dimensional conformal symmetry gets broken
by the interaction self-energy down to the conformal group $\mathrm{SL}(2,R)$ of rational transformations, $t\to
\tfrac{a t+b}{c t+d}$, $ad-bc=1$. This leads to a classic symmetry breaking scenario
and the emergence of Goldstone modes whose fluctuations become unhampered in the long
time limit where the \emph{explicit} symmetry breaking (represented by the time
derivative $\partial_t$ present in the system's action) becomes negligible.
The situation bears similarity to that in a magnet, with the important difference
that the dimension of the Goldstone mode manifold is infinite, while the spatial
dimension is zero. This means that Goldstone mode fluctuations are enhanced by the
Mermin-Wagner principle and expected to become virulent in the long time limit where the explicit symmetry breaking vanishes.

In Ref.~\cite{Bagrets:2016} we introduced a method to isolate these fluctuations and perform a full
integration over the Goldstone mode manifold. The idea is to introduce an
exponential reparameterization $f'(t)=\exp(\phi(t))$ whereupon the Goldstone mode
integral assumes the form of a path integral over $\phi$, with an (approximately)
time local action coinciding with that of so-called Liouville quantum mechanics \cite{Teschner:2001}.
Standard methods of quantum mechanics are then applicable to perform the
integration, including the regime where the $\phi$-fluctuations become large (but
those of a canonically conjugate `momentum', $k$, are small in exchange). We demonstrated
emergence of a new  time scale
\begin{align}
    \label{eq:M}
    M\equiv \frac{N\ln(N)}{64\sqrt \pi J}\, . 
\end{align}
At long times $t\gg M$,  Goldstone mode fluctuations are strong and act to \emph{restore} 
the broken symmetry (much like in a low dimensional magnet at weak external field rotational symmetry is
restored by unbounded fluctuations in the magnetization.) For example, while the mean-field 
Green function \cite{Sachdev:1993,Kitaev:2015,Maldacena:2016}, $G(\epsilon)\sim |\epsilon|^{-1/2}$ has a divergent amplitude
at low energies, the inclusion of Goldstone mode fluctuations shows \cite{Bagrets:2016} that
$G(\epsilon)\sim |\epsilon|^{1/2}$ for energies $|\epsilon|\ll M^{-1}$, i.e. symmetry
restoration supresses the mean-field propagator (the latter playing the role of an
average magnetization in the magnetic metaphor).

In Ref.~\cite{Bagrets:2016} we analyzed the effect of Goldstone modes within a zero temperature framework. However, the majority of observables describing the dynamical behavior of the SYK-model --- four point functions in general, and out-of-time-order (OTO) correlation functions in particular \cite{Maldacena1:2016, Larkin:1969, Maldacena2:2016} variants of the partition sum \cite{Sachdev:2015,Maldacena2:2016,Liu:2016}, etc. --- are formulated as quantum {\em thermal}  averages and require a finite temperature formalism. That this generalization is not entirely innocent is indicated by the observation that in the exponential degradation \cite{Maldacena2:2016,Maldacena1:2016}, $\sim\exp(2\pi Tt)$, of OTO correlations at small times, temperature, $T$, itself features as the relevant rate. 

Below we show that finite temperatures affect the effective quantum mechanics
describing the  Goldstone mode fluctuations via the appearance of an exponential
potential adding to the native potential of Liouvilian quantum mechanics. This
contribution strongly affects the Goldstone mode integral. To be specific, we consider the OTO correlation
function
\begin{align}
    \label{eq:OTO}
    F(t)\equiv \frac{1}{Z N^2}\sum_{ij}^N \left\langle\mathrm{tr}\left(\,e^{-\tfrac{\beta}{4}\hat H}\,\chi_i(0) \,e^{-\tfrac{\beta}{4}\hat H}\,\chi_j(t) \,e^{-\tfrac{\beta}{4}\hat H}\,\chi_i(0) \,e^{-\tfrac{\beta}{4}\hat H}\,\chi_j(t)\right) \right\rangle,
\end{align}
where $Z=\left\langle\mathrm{tr}\left(\,e^{-\beta\hat H}\right) \right\rangle$ is the partition sum. 
This expression, which differs from the standard definition of a thermal correlation function  by the symmetrization of thermal weights \cite{Maldacena2:2016,Maldacena1:2016}, has become an established tool for the diagnostics of  correlations in the model. 

\begin{figure}[t]
\centering{\includegraphics[width=14.0cm]{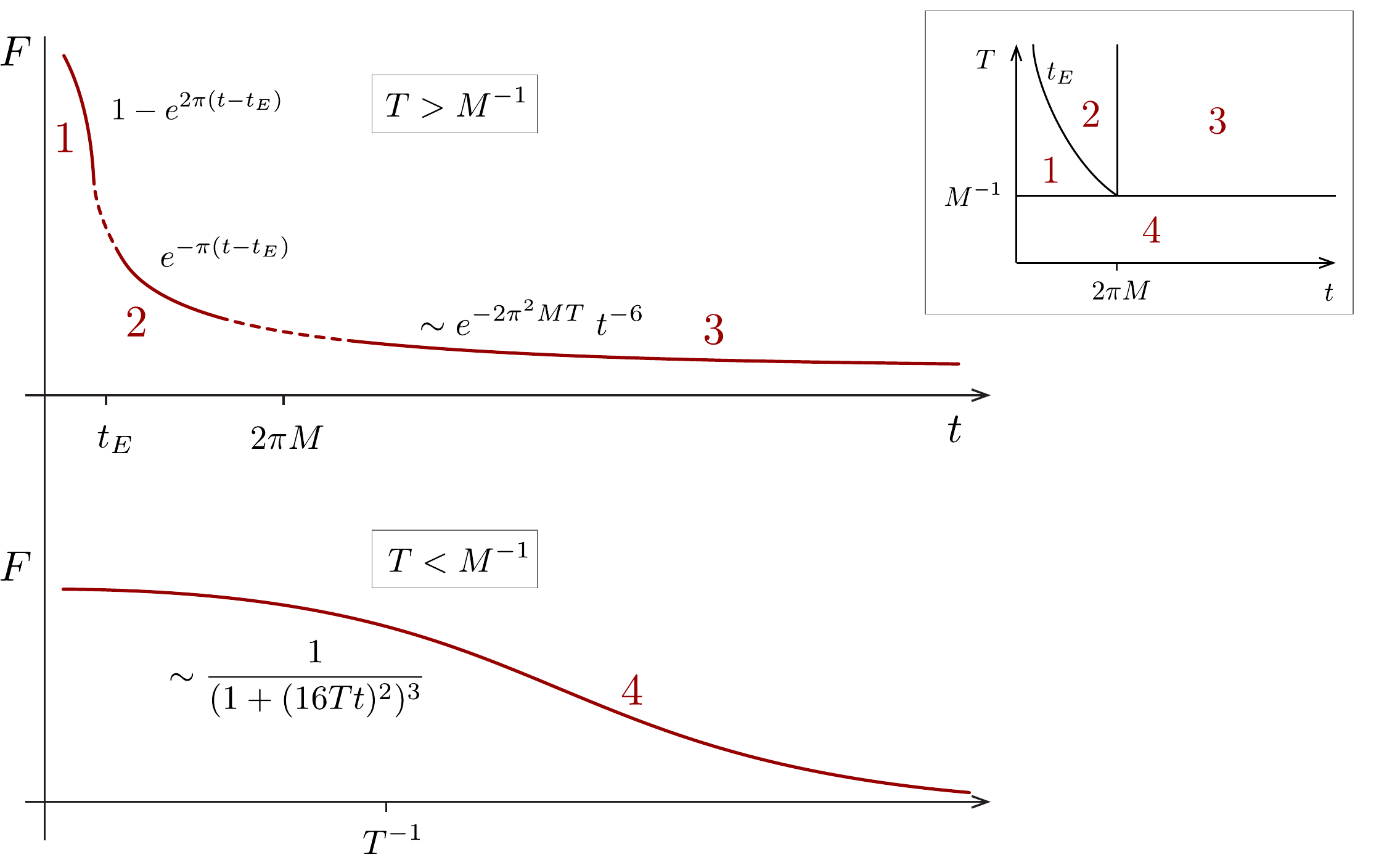}}
\caption{Results for the OTO correlation function. Top: At high temperatures, $T>M^{-1}$ and large times, $t>2\pi M$, the function crosses over from exponential to power-law decay with an exponent $t^{-6}$. Bottom: at low temperatures, $T<M^{-1}$ the function is nowhere exponential. At large times $t>T^{-1}>M^{-1}$ it again shows  $t^{-6}$ power-law behavior. The inset shows the parametric extension of the four regimes in a $t-T$ plane.}
\label{fig:OTOResults} 
\end{figure}

While previous work identified regimes of exponential decay of  OTO correlations
functions, here we show that Goldstone mode fluctuations are responsible for the formation of \emph{power laws}
at large times and/or low temperatures. Our main findings are summarized as follows. For high
temperatures, $T\gg M^{-1}$, we need to discriminate between short and large times, $t\ll
M$ and $t\gg M$, respectively. In the short time regime, Goldstone mode fluctuations
are weak, and one identifies \cite{Maldacena1:2016,Maldacena2:2016} two different regimes characterized by the exponential
loss of correlations, depending on whether $t<t_E$, or $t>t_E$, where
\begin{align}
\label{eq:Ehrenfest}
    t_E\equiv \frac{\ln(MT)}{2\pi T}
\end{align}
plays the role of an Ehrenfest time \cite{Larkin:1969} in the problem. However, at large
times, $t\gg M$, fluctuations are strong and generate universal power law scaling, $F(t)\sim t^{-6}$. For small temperatures, $T\ll M^{-1}$, Goldstone mode fluctuations affect the picture throughout the entire domain, no exponential regimes are found, and the power law, $F(t)\sim t^{-6}$, holds for all times $t\gg T^{-1}$. The quantitative summary of these statements reads as (cf. Fig.~\ref{fig:OTOResults} where the four distinctive regimes are indicated as $1,2,3,4$, respectively.)
\begin{align}
                   \label{eq:Summary}
T\gg M^{-1}:\qquad &F(t) = \left\{ \begin{array}{ll} 
1-\frac{1}{64\pi}\,e^{+2\pi T (t-t_E)}; \quad\quad& t< t_E\\
&\\
\ln(MT) \,e^{-\pi T (t-t_E)}; & t_E<t<2\pi M\\
&\\
(MT)^{-3/2}e^{-2\pi^2 M T } \left(\frac{M}{t}\right)^{6}; & 2\pi M< t
\end{array} \right.,\cr
T\ll M^{-1}:\qquad & F(t)=\frac{(MT)^{1/2}}{(1+(16 tT) ^2)^{3}},
\end{align}
where the algebraic profiles extend previously obtained results with exponential
behavior \cite{Maldacena:2016} to the regime of long times/low temperatures. 
These correlations appear as a robust consequence of the
Liouville quantum mechanics which effectively governs the long time behavior of the
system. Though we used 4-fermion model, Eq.~(\ref{eq:Hamiltonian}), as an example - 
the qualitative features, including $t^{-6}$ law, hold for an arbitrary number $q\geq 4$ of fermions in the interaction term.
This universality is based  on the $3/2$ power-law decay of two-time-points functions in the Liouville model \cite{Shelton:1998,Bagrets:2016}.   

The derivation of the above correlation laws also implies independent  validation of various results that have been obtained before on more phenomenological grounds (including by reference to principles of holography), or by different analytic methods. Notably, we find that the partition sum at small temperatures $T \ll J$ scales as
\begin{align}
    \label{eq:Z_sem}
    Z(\beta)\sim \left(  M/\beta \right)^{3/2}\exp(2\pi^2 M/\beta),\qquad \beta= \frac 1 T 
\end{align}
in agreement with the results of Ref.~\cite{Cotler:2016}. This formula is directly related to the 
many-body density of states (DoS), which we find for energies $E\ll J $ behaves as
\begin{equation}
\label{eq:DoS}
\rho(E) \propto \theta(E) \sinh\left[2\pi \sqrt{ 2 M E}\right],
\end{equation}
where the energy is counted from the ground state. 
This result was obtained previously in the limit of large number of interacting fermions, $q\sim \sqrt{N}$,  \cite{Cotler:2016} and by statistical analysis of moments of the random interaction operator \cite{Garcia:2017}. In our present analysis, the DoS reflects the energy stored in Goldstone mode fluctuations. 
We also note that at high temperatures,  $T \gg 1/M$, or energies, $E \gg 1/M$, the constant $M$ in Eqs.~(\ref{eq:Z_sem}) and (\ref{eq:DoS}) acquires a week logarithmic dependence on either
$T$ or $E$, respectively (see more comments on this in section~\ref{sec:SoftModeIntegration}).

Interestingly, the result~(\ref{eq:Z_sem}) for the partition sum entails an heuristic explanation for the power-law formation in the OTO correlation function~\eqref{eq:Summary} at long times/low temperatures, $t,T^{-1}<M$. To see this, let us insert four spectral resolutions of (realization specific) many body eigenstates $|m\rangle$ in~\eqref{eq:OTO} to obtain the `Lehmann representation'
\begin{align}
    \label{eq:OTOLehmann}
    F(t)= \frac{1}{Z N^2}\sum_{ij,m_i} \left\langle \langle m_1|\chi_i|m_2\rangle \langle m_2|\chi_j|m_3\rangle \langle m_3|\chi_i|m_4\rangle \langle m_4|\chi_j|m_1\rangle e^{-\left( \frac{\beta}{4}+it \right)\epsilon_{m_1}-\left(\frac{\beta}{4}-it \right)\epsilon_{m_2}  -\left(\frac{\beta}{4}+it \right)\epsilon_{m_3}-\left(\frac{\beta}{4}-it \right)\epsilon_{m_4}}\right\rangle.
\end{align}
Now, let us assume that the Majorana matrix elements $\langle m|\chi_{i,j}|n\rangle$ and the eigenenergies, $\epsilon_n$, are statistically independent. This assumption is of course grossly ad-hoc. However, statistical wave function/eigenvalue independence is a hallmark of random matrix models, and in view of the evidence for ergodic chaotic behavior shown by the SYK model at low energies \cite{You:2016,Garcia:2017}   may contain some truth. If so, and \emph{if} at the characteristic energy differences  $|\epsilon_m-\epsilon_m|\sim t^{-1}\sim M^{-1}$ much larger than the many-body level spacing $\sim e^{-N/2}$ statistical correlations between the levels of different spectral sums can be neglected, the  correlation function simplifies to  
\begin{align}
    \label{eq:OTOLehmannFactorized}
    F(t)\sim |Z(\beta/4+it)|^4,
\end{align}
where $Z(s)=\big\langle \sum_n  e^{-s \epsilon_n} \big\rangle=\int dE \,\rho(E)\, e^{- s E }$ with
${\rm Re}(s) >0$ is the Laplace transform of the average many-body DoS, Eq.~\eqref{eq:DoS},
which at $s=\beta$ is also identical to the physical partition sum. The large time behavior of such Laplace transform is dominated by branching points of the DoS function.  Since at energies $E<M^{-1}$ DoS exhibits non-analytic behavior as $\rho(E)\sim E^{1/2}$, it leads to $Z(s)\sim s^{-3/2}$ at $|s|\gg M$. This is exactly the 3/2 law, observed in the long-time behavior of two-point functions \cite{Bagrets:2016}. It in turn implies $F(t)\sim t^{-6}$ as announced in Eq.~\eqref{eq:Summary}. 

Of course, the construction above contains several ad-hoc assumptions and isn't trustworthy in its own right. The rest of the paper is devoted to a first principle validation of Eq.~\eqref{eq:Summary} from the low energy theory of SYK Goldstone mode fluctuations. We will start in section~\ref{sec:Preliminaries} where we  review the path integral description of  the model and the emergence of conformal soft modes. In section~\ref{sec:SoftModeIntegration} we construct the finite temperature Liouvillian soft mode action. In section~\ref{sec:SaddePoint} the short time fluctuation behavior of this effective theory will be studied by stationary phase methods on the example of the partition sum. Continuing with this quantity the analytical machinery required for studying strong Goldstone mode fluctuations are introduced in section~\ref{sec:PartitionQM}. The following four sections then define the core of the paper in which the OTO correlation function is addressed. We start by setting up a path integral representation of this quantity in section~\ref{sec:OTOPathIntegral}. Its short, intermediate, and long time behavior are then addressed in sections~\ref{sec:OTOStationary},~\ref{sec:OTOIntermediateTimes}, and~\ref{sec:OTOLongTimes}, respectively, before we conclude in section~\ref{sec:Discussion}. Several Appendices provide details on technical calculations.

\section{Preliminaries}
\label{sec:Preliminaries}
Our starting point is the effective action describing the system \cite{Sachdev:2015,Maldacena:2016} after the averaging over disorder and integration over Grassmann-fermion degrees of freedom have been performed (see~\ref{sec:StartingAction}) 
\begin{equation}
                        \label{eq:action}
- S[\Sigma,G]={N\over 2}\mathrm{Tr}\log(\partial_{\tau}\delta^{ab} + \Sigma^{ab}_{\tau,\tau'})  +
{N\over 2} \int\!\!\!\int\limits_{-\beta/2}^{\beta/2} d\tau d\tau' \left[   
 \frac{J^2}{4} \big(G^{ab}_{\tau,\tau'}\big)^4 +  \Sigma^{ba}_{\tau',\tau} G^{ab}_{\tau,\tau'}\right]. 
\end{equation} 
Here,  $G^{ab}_{\tau,\tau'}$ and $\Sigma^{ab}_{\tau,\tau'}$ are matrix fields
carrying an index structure, $a,b=1,\dots,n$ in a space of $n$ replicas. The first
term of the action resembles the free energy of a fermion system and identifies
$\Sigma$ as a  dynamical self energy. The second term reflects that the
average over disorder, $\langle\hat H^2\rangle$ is of eighth order in Majorana
operators. This is amounts to four Green functions, which here play a role of
dynamical matrix fields, too. The third term expresses the conjugacy of the two
fields $\Sigma,G$. 

The action~\eqref{eq:action} exhibits an exceptionally high level of symmetry, provided $\partial_\tau$ term is neglected: it is
then invariant under reparameterizations of time, $\tau \to f(\tau)$, where $f(\tau)$ may
be any invertible and  differentiable function. (Invertibility requires globally
monotonicity and without loss of generality we assume $f'(\tau)>0$ throughout.  The
condition of differentiability may be sacrificed at isolated points if necessary.)
This means that $S$ possesses an infinite dimensional symmetry group whose generators
are the Virasoro generators of infinitesimal reparameterization transformations,
$f$.

The interpretations of the fields $G$ and $\Sigma$ become more tangible at the mean-field level, where the stationary phase equations~\cite{Sachdev:2015} assume the form of the  self-consistent  Dyson equation 
\begin{eqnarray}
              \label{eq:Dyson}
 - (\partial_{\tau} + \Sigma)\cdot G = 1; \quad \quad \Sigma=  J^2 \big[G\big]^3.
\end{eqnarray}
The first equation here is a matrix equation while in the second the cube operation
acts on each matrix element separately, i.e. $[G]^3\equiv \big[G^{ab}_{ \tau,
\tau'}\big]^3$.  These equations can be solved in the long time limit $\tau-\tau'\gg
1/J$, where $\partial_{\tau}$ may be neglected. A configuration,
$G_\infty,\Sigma_\infty$ solving the equations at zero temperature, $\beta\to
\infty$, reads as
\begin{equation}
               \label{eq:bare-solution}
G^{ab}_\infty(\tau-\tau')= - \frac{b}{J^{1/2}}\,\frac{\delta^{ab}\,\mathrm{sgn}(\tau-\tau')}{|\tau-\tau'|^{1/2}}; \quad \quad
\Sigma^{ab}_\infty(\tau-\tau') =   - b^3J^{1/2}\,  \frac{\delta^{ab}\,\mathrm{sgn}(\tau-\tau')}{|\tau-\tau'|^{3/2}},
\end{equation}
where $b=(4\pi)^{-1/4}$. However, this solution is not unique
\cite{Kitaev:2015,Polchinski:2016,Maldacena:2016}. Under the action of the symmetry
operation it transforms as
\begin{align}
               \label{eq:soft-manifold}
G_\infty([f],\tau,\tau') &= f'(\tau)^{1/4}G_\infty\bigl(f(\tau)-f(\tau')\bigr) f'(\tau')^{1/4}=- \frac{b}{J^{1/2}}\mathrm{sgn}(\tau-\tau')\,\frac{f'(\tau)^{1/4} f'(\tau')^{1/4}}{|f(\tau)-f(\tau')|^{1/2}}, \cr
\Sigma_\infty([f],\tau,\tau') &=  f'(\tau)^{3/4}\Sigma_\infty\bigl(f(\tau)-f(\tau')\bigr) f'(\tau')^{3/4}=   - b^3J^{1/2}\mathrm{sgn}(\tau-\tau')\,  \frac{f'(\tau)^{3/4} f'(\tau')^{3/4} }{|f(\tau)-f(\tau')|^{3/2}},
\end{align}
i.e. we encounter the classic scenario where a symmetry is broken at the mean field level, and a Goldstone mode manifold emerges. One may verify that the sole transformations leaving the saddle points invariant are the conformal reparameterizations, $\tau \to \tfrac{a\tau+b}{c\tau +d}, ad-bc=1$,  i.e. the Goldstone  mode manifold is identified as the group of time-reparameterizations modulo the unbroken $\mathrm{SL}(2,R)$-transformations. 

The transformation (\ref{eq:soft-manifold}) provides a way to  generalize the saddle-point solution from one defined on the entire real axis to a finite-temperature solution with support on $[-\beta/2,\beta/2]$  \cite{Georges:1999}:
the reparameterization
\begin{equation}
\label{eq:g-function}
 \tau \rightarrow g(\tau) := \tan \bigl( \pi \tau/\beta \bigr), 
\end{equation}
provides an invertible map from the open imaginary time interval $[-\beta/2,\beta/2]$ onto the entire axis.  According to Eq.~(\ref{eq:soft-manifold}) the corresponding Green function reads
\begin{equation}
\label{eq:G-beta}
G_\beta(\tau - \tau') = G_\infty([g],\tau-\tau')= 
- \frac{b}{J^{1/2}} \mathrm{sgn}(\tau-\tau') 
\left[\frac{\pi}{\beta \sin ({\pi |\tau-\tau'|}/{\beta})} \right]^{1/2}. 
\end{equation}
and similarly for $\Sigma_\beta(\tau_1-\tau_2)$. 
This defines a periodic and time translationally invariant solution of the finite temperature saddle-point equation (\ref{eq:Dyson}). The group property of the symmetry implies that reparameterization of compactified time $\tau\to f(\tau)$, mapping the interval $[-\beta/2,\beta/2]$ onto itself and obeying the periodicity constraint
\begin{equation}
\label{eq:periodicity}
f(\tau+\beta)-f(\tau)=\beta; \quad\quad  f'(\tau+\beta)=f'(\tau)\geq 0.
\end{equation} 
generate a family of finite temperature solutions 
$G_\beta([f],\tau, \tau')$, periodic on the time interval but lacking translational invariance (i.e. the functions $G_\beta[f]$ depend two time arguments separately and not just the difference).

\section{Soft mode integration}
\label{sec:SoftModeIntegration}

The low energy properties of the model are described by a  functional integral over
the soft-mode manifold parameterized by the functions $f$ identified in the previous
section. In Ref.~\cite{Bagrets:2016} we showed that at zero temperature the emerging
integrals can be understood as path integrals of Liouville quantum mechanics. In the
following, we discuss the   non-trivial generalization to finite
temperatures to then  apply it to the computation of observables.

\begin{figure}[t]
\centering{\includegraphics[width=13cm]{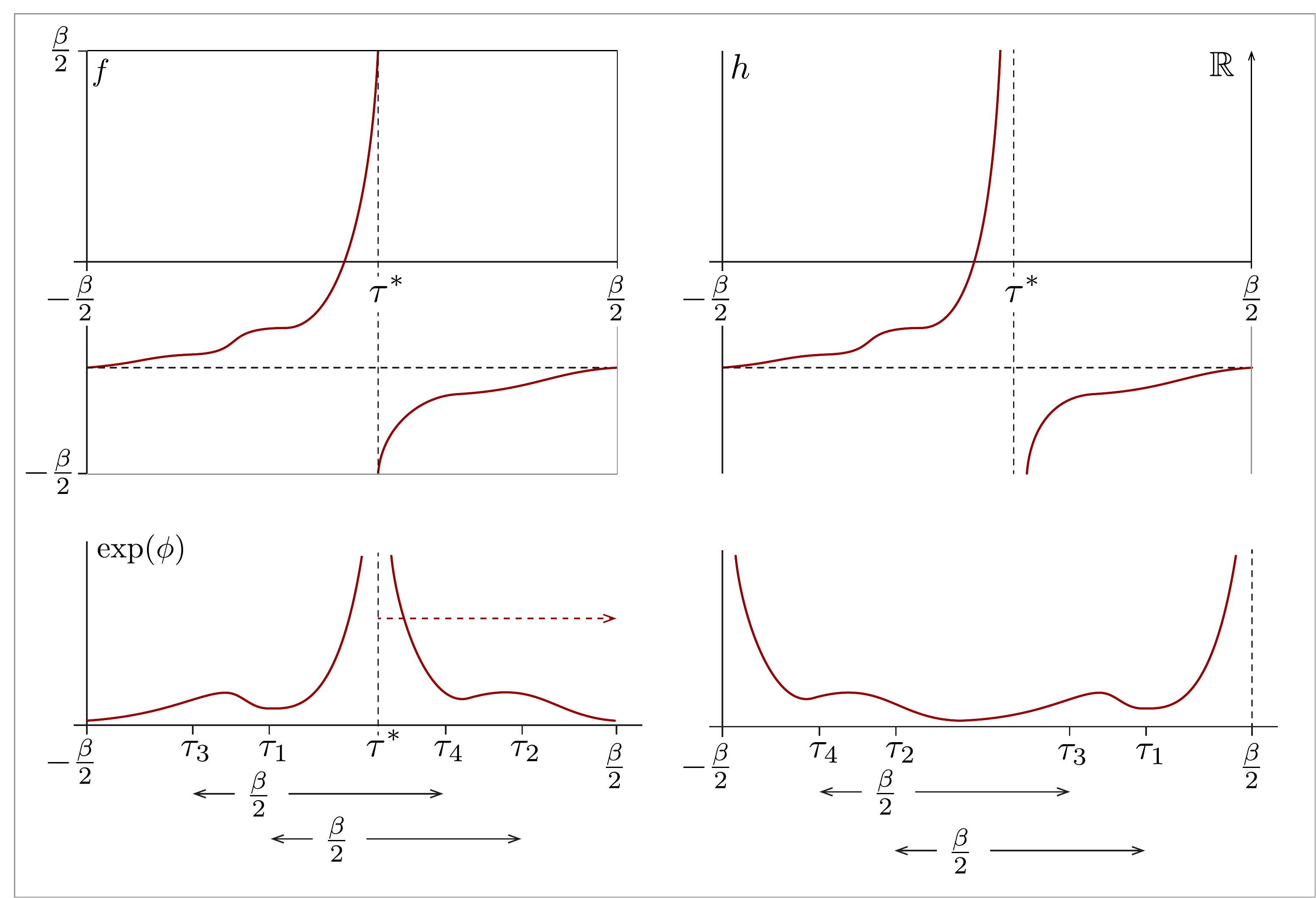}}
\caption{Top: an invertible map $f(\tau)$ from the interval $[-\beta/2,\beta/2]$ onto itself (left) and $h(\tau) \equiv g(f(\tau))$ 
from the interval $[-\beta/2,\beta/2]$  onto the reals (right). In the latter case a singularity at some $\tau^\ast$ is necessarily present. Bottom left:
reparameterization of $h'=\exp(\phi)$. For later reference, the imaginary time
arguments $\tau_1,\dots,\tau_4$ of the OTO correlation function are indicated. Bottom
right: a shift $s^\ast \to \beta/2$ is applied to move the singularity to the
boundaries of the time interval.}
\label{fig:Reparameterization}
\end{figure}

Rather than integrating over functions $f:[-\beta/2,\beta/2]\to [-\beta/2,\beta/2]$
mapping the finite imaginary time interval onto itself (cf.
Eq.~\eqref{eq:periodicity}), we find it convenient to work with functions
$h:[-\beta/2,\beta/2]\to \Bbb{R}$ which have the full real numbers as their image.
(Within the $[-\beta/2,\beta/2]\to [-\beta/2,\beta/2]$ framework, the weak symmetry
breaking part of the effective action picks up potential contributions under
re-transformations which are unpleasant to work with.) The passage to this
description is defined by $h(\tau)\equiv g(f(\tau))$, where $g$ is the
$\tan$-function, Eq.~(\ref{eq:g-function}). The function $h$ then necessarily has a
singularity at some $\tau^*\in  [-\beta/2,\beta/2]$, where $f(\tau^*)=\pm \beta/2$
(cf. Fig.~\ref{fig:Reparameterization}.) A shift of the time axis, which is part of
the $\mathrm{SL}(2,R)$ invariance manifold, can be applied  to shift  the singularity
to the boundaries of the interval $\tau^*=\pm\beta/2$.

Recalling that $G_\beta=G_\infty([g])$, we have $G_\beta[f])=G_\infty([g\circ f])=G_\infty[h]$, or 
\begin{equation}
G_\beta([f],\tau,\tau')
= h'(\tau)^{1/4} G_\infty\bigl( h(\tau) - h(\tau') \bigr) h'(\tau')^{1/4}, 
\label{eq:G_h}
\end{equation}
in a more explicit representation.  Notice that, although we effectively returned to
the zero temperature Green function, the time arguments lie in the finite temperature
interval $\tau,\tau'\in  [-\beta/2,\beta/2]$.

Likewise, any finite-temperature observable, ${\cal O}_\beta[f]$, formulated in terms
of reparameterizations, $f$, of the finite temperature  soft mode manifold may be
transformed into the {\em zero temperature} observable  $ {\cal O}_\infty[h]$,
parameterized by $h=g\circ f$. Expectation values of such observables are obtained
by integration over reparameterizations,
\begin{equation}
\label{eq:observable} 
\langle  {\cal O}_\beta \rangle \propto  \int \mu[f]\, {\cal D}f\,   {\cal O}_\beta[f]\,   e^{-S_\beta[f]} \, \overset{ f = g^{-1} \circ h}{ = }\,
\int \mu[h]\, {\cal D}h\,   {\cal O}[h]\,  e^{- S[h]} ,   
\end{equation} 
where we omitted the subscript $'\infty'$ for brevity. Here, $\mu[ h ] \propto
{\prod_\tau} [h'(\tau)]^{-1}$~\cite{Bagrets:2016} is the group measure whose reparameterization invariance  implies $\mu[f]\mathcal{D}f=\mu[h]\mathcal{D}h$.
The action $S[h]$ in Eq.~(\ref{eq:observable}) describes the cost of $h$-fluctuations due to the presence of the derivative  $\partial_\tau$ in Eq.~(\ref{eq:action}). Requiring invariance under the $\mathrm{SL}(2,R)$-reparameterization group and {\em assuming} locality in time, one may argue \cite{Maldacena:2016} that this action must be the integrated Schwarzian derivative     
\begin{equation}
\label{eq:h-action}
 S[h]= -M \int\limits_{-\beta/2}^{\beta/2} d\tau\, \{h(\tau),\tau\}, 
\end{equation}
where $\{h,\tau\} \equiv ( \tfrac{h''}{h'})'- \tfrac 12 ( \tfrac{h"}{h'})^2$  and $M$ is a coupling constant of dimensionality $[\mathrm{time}]$\footnote{
The action $S_\beta[f]$ is obtained from here as $S_\beta[f] \equiv S[g \circ f] = S[f] - \frac{2 \pi^2 M}{\beta^2}
\int\limits_{-\beta/2}^{\beta/2} d\tau\, f'^2$, cf. Ref~\cite{Maldacena:2016}. }. 
On dimensional grounds, this constant determines the time scale, $t_* \sim M$, above which the reparameterization fluctuations become strong. 
The value of this constant can be fixed by a more explicit construction~\cite{Bagrets:2016} which first translates the starting action~\eqref{eq:action} into a \emph{non-local} $f$-action and then notes that the collapse to an effectively local action takes place below a certain energy scale which has to be determined by a self consistency-analysis~\cite{Bagrets:2016}. 

In more concrete terms, one defines $M$ as the running coupling constant of the Schwar\-zian theory, $M(E)$, which has a weak logarithmic dependence on a typical energy involved 
when the latter is sufficiently high, $E > M(E)$, namely
\begin{equation}
M(E) = \frac{N \ln(J/E)}{64 \sqrt{\pi} J} ,  \qquad E > 1/M(E).
\end{equation}
This logarithmic renormalization stops at the scale $\Delta$ defined self-consistently via $\Delta  = 1/M(\Delta)$, which with log accuracy is resolved as $\Delta = 64 \sqrt{\pi} J/ (N \ln N)$.
At smaller energies, $E < \Delta$, the Goldstone mode fluctuations proliferate and $M(E)$ saturates to the constant value $M \equiv 1/\Delta$ 
as stated in equation~\eqref{eq:M}. Note, that a logarithmic dependence of $M(E)$ is the price to pay to reduce the intrinsically non-local action for reparametrizations to the local Schwarzian form. 
In practical calculations one should take $M$ at the scale $E = {\rm min}(T, 1/t)$, where $t$ is a time variable in the correlation function. 

Before turning to the actual integration procedure in Eq.~(\ref{eq:observable}), we impose one more change of variables and introduce the degree of freedom, $\phi(\tau)$, of Liouville quantum mechanics as
\begin{equation}
h'(\tau) \equiv e^{\phi(\tau)},\quad \quad \phi(\tau+\beta)=\phi(\tau),
\label{eq:phi_h}
\end{equation}
where the positivity $h'(\tau)>0$ guarantees that  $\phi(\tau)$ is real.
The key advantage of this parameterization is the flatness of its
integration measure, $\mu[h]\, {\cal D}h ={\cal D}\phi$~\cite{Bagrets:2016}. However, attention must be
payed to the singularity of all admissible $h$-functions at $\tau=\pm\beta/2$. It
translates into a diverging integral $\int_{-\beta/2}^{\beta/2} d\tau\, h'(\tau) =
\int_{-\beta/2}^{\beta/2} d\tau\, e^{\phi(\tau)}=\infty$. We find it convenient to
regularize this divergence by introducing a finite but long time $t_H$ such that
\begin{equation}
\label{eq:t-H}
\int\limits_{-\beta/2}^{\beta/2} e^{\phi(\tau)} d\tau = t_H,
\end{equation}
and take a limit $t_H\to \infty$ at the end of the calculation. 
Trajectories $\phi(\tau)$ satisfying this condition thus become regularized at 
the boundaries of the time interval, $\phi(\pm\beta/2) = \phi_0$ with $\phi_0 \to
+\infty$ as $t_H \to \infty$. 

In the $\phi$-language the soft mode action reads as
\begin{equation}
\label{eq:h-action_phi}
 S[h]\to S[\phi]={M\over 2}\int\limits_{-\beta/2}^{\beta/2} d\tau \left[ (\phi')^2 - 2\phi''\right]. 
\end{equation}
and expectation values of  observables ${\cal O}_\beta$ assume the form 
\begin{equation}
\label{eq:observable1} 
\langle  {\cal O}_\beta \rangle \propto \, {Z_\beta}^{-1}\!\!\!\!\!
\int\limits_{\phi(-\beta/2)=\phi_0}^{\phi(\beta/2)=\phi_0} {\cal D}\phi\,   {\cal O}[\phi]\,  e^{-S[\phi] } ,\qquad S[\phi]= \int\limits_{-\beta/2}^{\beta/2} d\tau\, \left\{ {M\over 2}\left[ (\phi')^2 - 2\phi''\right] + \gamma \, e^{\phi(\tau)}  \right\}-\gamma t_H, 
\end{equation} 
where $\gamma $ is a Lagrange multiplier  enforcing the condition (\ref{eq:t-H}),
and $Z_\beta=\int \mathcal{D}\phi\exp(-S[\phi])$ is the partition sum.
Notice that the full derivative $\phi''$ cannot be ignored since, due to singularity
discussed above, $\phi'(\pm\beta/2)\to \pm\infty$. It is this term which neutralizes the
formally infinite constant $\gamma  t_H$. The same singularity also shows in the boundary
conditions  $\phi(\pm\beta/2) = \phi_0$. Our treatment below establishes a balance between the   parameters
$t_H$,  $\phi_0$, and $\gamma $ such that the final answers do not depend on regularization specifics.

\section{Saddle point calculation}  
\label{sec:SaddePoint}

At sufficiently high temperature, and/or sufficiently short time intervals $\tau-\tau'$ the functional integral (\ref{eq:observable1}) may be evaluated in a saddle point approximation. This saddle point approach is stabilized by the largeness of $M$ compared to all other time scales in the problem. To expose this parameter in the best possible way we introduce a rescaled variable as
\begin{align}
    \label{eq:VarphiScaling}
    \phi\equiv \ln \left( \frac{2M}{\gamma} \right) + \varphi,
\end{align}
where we are simplifying the notation by temporarily setting $\beta\equiv \pi$. In the final results the $\beta$-dependence can then be re-introduced by scaling all quantities with dimension of time as $\tau\to \tau \pi/\beta$. Expressed in the new language, the action assumes the form
\begin{align}
    \label{eq:ActionScaled}
    S[\varphi]=M\int\limits_{-\pi/2}^{\pi/2} d \tau \, \left(\frac{1}{2} [\varphi'(\tau)]^2 +  2  e^{\varphi(\tau)}\right),
\end{align}
the constraint becomes
\begin{align}
    \label{eq:tHVar}
    \int\limits_{-\pi/2}^{\pi/2}  e^{\varphi(\tau)}d \tau=\frac{ t_H \gamma}{2M},
\end{align}
and the equation of motion reads as $\varphi''(\tau) = 2  e^{\varphi(\tau)}$. In the limit $t_H \to +\infty$ this is solved by  $ \bar\varphi_\infty(\tau) = 
 - \ln\left ( \cos^2\tau\right)$.
(The dependence on $t_H$ is implicit through Eq.~\eqref{eq:tHVar}, which for $t_H\to \infty$ implies diverging boundary conditions $\bar\varphi_\infty(\tau\to \pm \pi/2)\to \infty$.)
The meaning of this solution becomes apparent when we formulate it in the language of
the $h$-representation, $\bar h_0(\tau)=\frac{2M}{\gamma}\int d\tau e^{\bar\varphi_\infty(\tau)}=\frac{2M}{\gamma}
g(\tau)$, where $g$ is the $\tan$-function of Eq.~\eqref{eq:g-function}. It is straightforward to check that for this function $G_\infty([\bar
h_0],\tau,\tau')=G_\beta(\tau-\tau')$ reduces to the  familiar \cite{Georges:1999} translationally
invariant finite temperature solution of the Dyson equation (\ref{eq:G-beta}).

Finite values of  $t_H$ regularize the infinity in the boundary conditions of
$\bar\varphi_\infty$. This regularization can be achieved by  shifting the argument of the $\cos$ in the solution as,
\begin{align}
    \label{eq:phi_opt}
    \bar \varphi(\tau)\equiv  \ln\frac{(1-\lambda)^2}{\cos^2\Big(\tau(1-\lambda)\Big)}, \qquad \lambda \ll 1.
\end{align}
For small $\lambda$ we find that the integral in~\eqref{eq:tHVar} evaluates to $4/(\pi\lambda)$, and this fixes the shift parameter as
\begin{align}
    \label{eq:EpsilonParameter}
    \lambda=\frac{8M}{\pi \gamma t_H}.
\end{align}
For this configuration, we find a boundary value, $\exp(\bar\varphi(\pm \pi/2))\equiv  \exp(\varphi_0)=(2/\pi\lambda)^2$. 

Now is a good time to discuss the status of the Lagrange multiplier, $\gamma$. This
parameter was introduced to enforce the constraint, Eq.~\eqref{eq:tHVar}. In
conventional variational calculus, the value of a Lagrange multiplier is fixed by
requiring that the constraint is satisfied on the variational configuration $(\bar
\varphi)$. Presently, however, we have one more free variable in the picture, viz.
the diverging boundary values $\varphi_0$. We may use this freedom to make sure that
for \emph{any} given value of $\gamma$  the constraint is satisfied (we saw above
that this is achieved by the value $\exp(\varphi_0)=(2/\pi\lambda)^2=(\gamma
t_H/4M)^2$.) This gives us the freedom to fix $\gamma$ at will, and the best choice
is $\gamma=J$, i.e. the largest possible value of a variable with dimension
$[\mathrm{energy}]$ in the problem. A large value of the Lagrange multiplier is
favored because it minimizes fluctuation-violations of the constraint (much like
particle number fluctuations in statistical mechanics are minimized by large chemical
potentials. For a few more remarks substantiating this point,
see~\ref{sec:LagrangeRemarks}.) Our rationale, thus, is to impose a fixation
$\gamma=J$ throughout. This determines the boundary values of the Liouville quantum
mechanics which in the language of the original variable, $\phi$  read
\begin{align}
    \label{eq:PhiBoundaryCondition}
    \exp(\phi_0)=\exp(\phi(\pm \pi/2))=\frac{\gamma t_H^2}{8M}.
\end{align}
Finally, a straightforward calculation shows that the stationary phase action is given by 
\begin{equation}
S[\bar\varphi]=M\int\limits_{-\pi/2}^{\pi/2} d\tau\, \left\{\frac{1}{2} (\bar\varphi')^2 - \bar\varphi'' + 2 \, e^{\bar\varphi(\tau)}  \right\} - \gamma t_H = -2\pi M,
\label{eq:F_h0}
\end{equation}
independent of the regularization. 

Gaussian fluctuations around $\bar\varphi$ lead to a fluctuation determinant whose  calculation is detailed in~\ref{sec:Gaussian}. Multiplying this factor with the exponentiated saddle point action~\eqref{eq:F_h0}, leads to the result~\eqref{eq:Z_sem} scaled by the regularization-dependent prefactor $(\gamma t_H)^{-1}$. The latter may be considered as an artifact of the path integral normalization and drops out in all physical observables normalized by the partition sum. 
 
\section{Partition function from Liouville quantum mechanics}
\label{sec:PartitionQM}

We next discuss  different approach to computing observables which is tailored to handle regimes of  large fluctuations. The idea is to map the Liouville path integral  to an equivalent Schr\"odinger equation and solve the latter.
Where the partition function is concerned, this procedure leads to results identical
to those of the stationary phase approach. The exactness of the latter in connection
with  $Z$ has indeed been stated before \cite{Cotler:2016}, although we are not aware of a proof and the
partition sum does not appear to satisfy  standard criteria \cite{Szabo:2000}  for the
semiclassical exactness of path integrals in an obvious ways. At any rate, pronounced
deviations between the two approaches will be observed when other quantities are considered.

The functional integral in Eq.~(\ref{eq:observable1}) lends itself to a straightforward  quantum mechanical interpretation. To this end,  consider the  Liouville Hamiltonian operator
\begin{equation}
\label{eq:Hamiltonian_QM}
\hat H = - \frac{\partial_\phi^2}{2M} + \gamma  e^\phi,
\end{equation}
and the corresponding imaginary time  propagator:
\begin{equation}
\label{eq:K_QM}
\Pi(\phi_1,\phi_2,\tau) = \langle \phi_1 | e^{-\tau \hat H} | \phi_2 \rangle = 
\int\limits_{\phi(-\tau/2)=\phi_1}^{\phi(\tau/2) = \phi_2}   
{\cal D}\phi \,\, e^{-\int\limits_{-\tau/2}^{\tau/2} d s \, \left\{ {\frac M 2} [\phi'(s)]^2 + \gamma   e^{\phi(s)}\right\}}.
\qquad 
\end{equation}
Comparison with Eq.~(\ref{eq:observable1}) shows that the regularized Schwarzian partition function can be identified with the quantum propagator
\begin{equation}
\label{eq:ZvsPi}
 Z(\beta) =    e^{2\gamma t_H}\,\Pi(\phi_0,\phi_0,\beta)\, . 
\end{equation} 
\emph{provided} we have the identification 
\begin{equation}
\label{eq:gamma-tH}
M \int\limits_{-\beta/2}^{\beta/2} d s\, \phi''(s)  = M\phi'(s)\stackrel{!}{=} M  \bar\phi'(s) \Bigl |_{-\beta/2}^{\beta/2} = \gamma  t_H.
\end{equation}
We here require that at the integration boundaries, $\phi(\pm\beta/2)\simeq
\bar\phi(\pm\beta/2)$ the semiclassical approximation remains valid, including in
regimes where fluctuations are otherwise large. The justification for this assertion
is that in the limit $t_H\to \infty$ the boundary field amplitudes assume large
values and fluctuations are relatively less pronounced than in the interior of the
integration interval.  Under these conditions, the semiclassical expressions
(\ref{eq:phi_opt}), (\ref{eq:EpsilonParameter}) yield Eq.~(\ref{eq:gamma-tH}), which in turn leads to the equivalence
$\Pi\sim Z$.

The Hamiltonian $\hat H$, Eq.~(\ref{eq:Hamiltonian_QM}), has a continuous spectrum
labeled by a quantum number $k$ which may be interpreted as the  momentum conjugate
to $\phi$. Its energy eigenvalues are given by $E_k = k^2/2M$, and are independent of $\gamma$.  The dependence on this parameter is in the normalized
wave functions which take the form
\begin{equation}
 \label{eq:eigenfunction}
\langle \phi  |k\rangle = \Psi_k(\phi)={\cal N}_k K_{2ik}\left(2\sqrt{2 M\gamma}\, e^{\phi/2} \right), \quad\quad\quad {\cal N}_k= \frac{2}{\Gamma( 2 i k)},
\end{equation}
where $K_\alpha$ are modified Bessel functions, and $\Gamma$ is the Gamma function. 
With these functions the spectral representation of the  propagator (\ref{eq:K_QM}) reads as
\begin{equation}
\Pi(\phi_1,\phi_2,\tau)  = \int\limits_0^{+\infty} \frac{dk}{2\pi}\, \Psi_k(\phi_1) \Psi^*_k(\phi_2)\, e^{-k^2 \tau/2M}. 
\end{equation}
Into this expression we substitute $\phi_1=\phi_2=\phi_0$, which means that we have to consider the (real) wave function amplitudes
\begin{align}
    \Psi_k(\phi_0)={\cal N}_k K_{2ik}\left(2\sqrt{2 M\gamma}\, e^{\phi_0/2} \right)\stackrel{\eqref{eq:PhiBoundaryCondition}}{=}\mathcal{N}_k K_{2ik}(\gamma t_H)\sim \mathcal{N}_k e^{-\gamma t_H}/\sqrt{\gamma t_H},
\end{align}
where the large argument asymptotic of the Bessel functions was used. Noting that 
 $(\Gamma(2ik)\Gamma(-2ik))^{-1}=2k\sinh(2\pi k )/\pi$ the partition function~\eqref{eq:ZvsPi} assumes the form
\begin{align}
\label{eq:ZkIntegral}
Z(\beta) \sim  \frac {1}{\gamma t_H} 
 \int\limits_{0}^{+\infty} dk k \sinh(2\pi k)  \,e^{-\beta k^2/2M} = \frac {M}{\gamma t_H}   \int\limits_0^\infty dE \, \sinh\left[2\pi\sqrt{2M E}\right] \, e^{-\beta E}.
\end{align}
The last expression implies that the average {\em many-body} DoS, $\rho(E)$, of the SYK model is given by~\eqref{eq:DoS}.
Here, $E$ measures the positive energy of reparameterization fluctuations and $\rho(E)$ is the corresponding density of states relative to the ground state energy.  This remarkably simple expression was obtained in Ref.~\cite{Cotler:2016} in the limit of a large number of interacting fermions and more recently in Ref.~\cite{Garcia:2017} by the combinatorial analysis of averaged moments of the Hamiltonian  operator.  We here show how it follows from the  Liouville partition function.

The square-root singularity of the DoS at low energies resembles the behavior of effective random matrix models close to their mean field spectral band gaps. In the case of SYK the corresponding square root behavior is observed at  the energy scale $\sim 1/M\sim J/N\log N$, where the reparameterization fluctuations become strong. As we explain in the Introduction it is this branch cut singularity of the many-body DoS which is responsible for the universal power-law behavior of the correlation functions at 
long times. 

Finally, performing  integration over $k$ in Eq.~\eqref{eq:ZkIntegral}, we obtain the partition sum~\eqref{eq:Z_sem} 
in exact  agreement with the Gaussian approximation.

\section{Out of time correlation function I: path integral representation} 
\label{sec:OTOPathIntegral}

\begin{figure}[t]
\centering{\includegraphics[width=12.0cm]{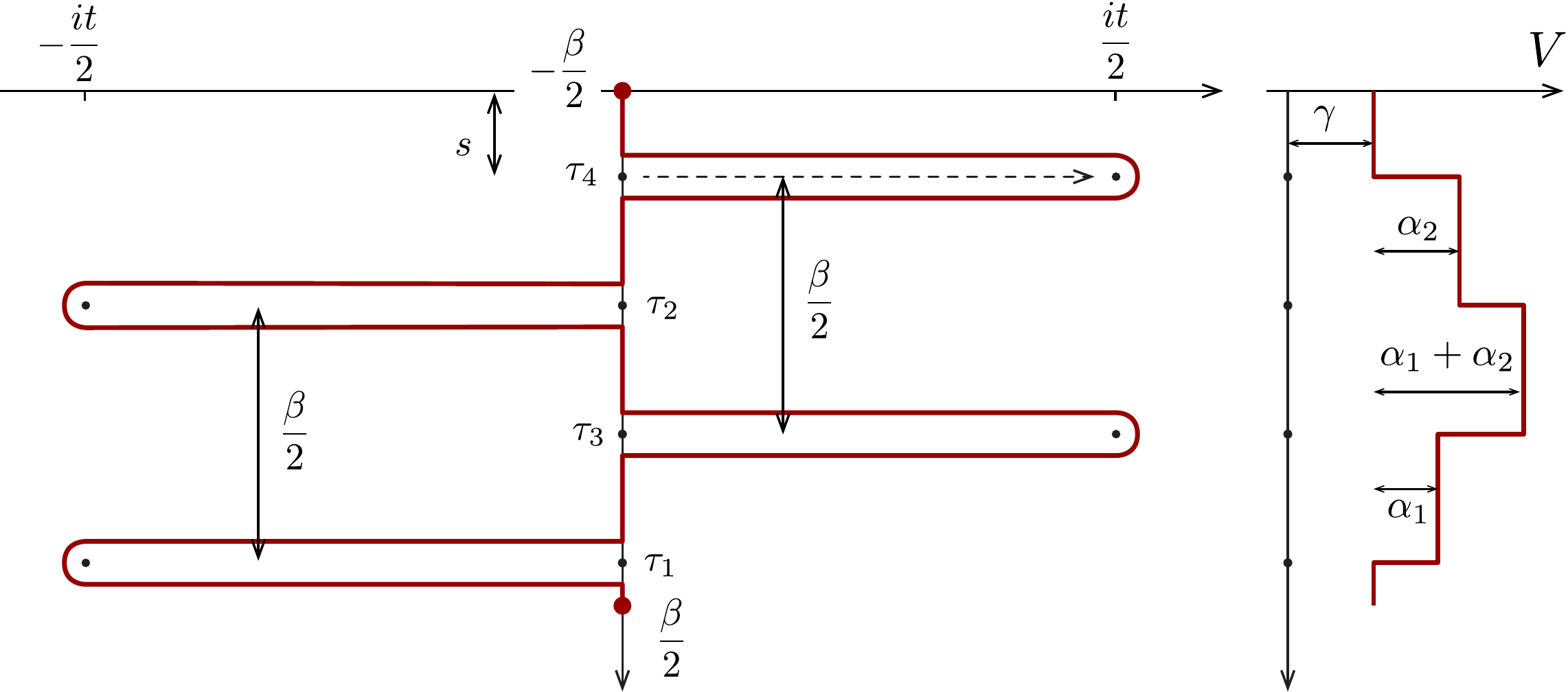}}
\caption{Time arguments entering our analysis of the OTO correlation function in the complex plane. Discussion, see text. The red line can be understood as a general path ordering prescription underlying the definition of path integral representations in the theory. The right panel shows the corresponding quenches of the Liouville potential as a function of the imaginary time.}
\label{fig:Contour}
\end{figure}

In this and the following sections we discuss the dynamics of the OTO correlation function.  Unlike with the partition sum, it will turn out that now the proper treatment of fluctuations beyond the Gaussian level becomes essential.
Let us consider the general four point function
\begin{align}
\label{eq:Green-4}
G_4(\tau_1,\tau_2,\tau_3,\tau_4) \equiv \frac{1}{N^2} \sum_{i,j}
\langle T_\tau \chi_i(\tau_1) \chi_i(\tau_2) \chi_j(\tau_3) \chi_j(\tau_4) \rangle,      
\end{align}  
where the angular brackets denote both, disorder averaging, and the quantum thermal average $\mathrm{tr}(\exp(-\beta \hat H)(\dots))$, and $T_\tau$ is imaginary time ordering. From this function, the OTO correlation function~\eqref{eq:OTO} is obtained by the specific choice of complex time arguments (see Fig.~\ref{fig:Contour})
\begin{align}
\tau_1 = \frac {\beta}4 +s -\frac{it}{2}, \quad \tau_3 = s+ \frac{it}{2}, \quad \tau_2 = -\frac\beta 4 +s -\frac{it}{2}, \quad \tau_4 = -\frac{\beta}{2}+s+\frac{it}{2}, 
\label{eq:C4Times}
 \end{align}
at $s=0$,  and the time ordering is along the contour \cite{Aleiner:2016} shown in Fig.~\ref{fig:Contour}.  Here
we used the fact that the correlation function depends only on differences between
neighboring times to switch to a symmetric arrangements of the arguments $\pm it/2$
relative to the real axis. (This configuration turns out to be more convenient in the
subsequent integration over fluctuations.) Throughout, we will largely  work with
real time arguments, $\tau_i\in [-\beta/2,\beta/2]$ and perform the analytic
continuation to finite imaginary increments $\pm it/2$  only at the final stages of
the calculations. The shift parameter, $s$, in Eq.~\eqref{eq:C4Times} enters the
stage when the position of the reparameterization singularity at $\tau^\ast$, Fig.~\ref{fig:Reparameterization}, relative
to the observation times becomes of importance. In such cases, we shift $\tau^\ast\to
-\beta/2$ (which is always possible on account of the periodicity of the imaginary
time theory) and offset the observation times by a parameter, $s$, which then is
integrated over.

We will work under the assumption that only soft reparameterization fluctuations are essential to the behavior of this function.
The four point function then assumes the form of a product of two two point functions,
\begin{align}
    G_4(\tau_1,\tau_2,\tau_3,\tau_4) =\langle  G_\beta([f],\tau_1,\tau_2)  G_\beta([f],\tau_3,\tau_4) \rangle,
\end{align}
where $G_\beta([f])$ is the reparameterized Green function~\eqref{eq:G_h} and the functional average defined by Eq.~\eqref{eq:observable1} leads to correlations between them. To bring the Green functions into a form suitable for  functional integration, we substitute~\eqref{eq:phi_h} into~\eqref{eq:G_h} and use~\eqref{eq:bare-solution} to obtain
\begin{align}
    \label{eq:GreenFunctionAlpha}
    G_\beta([f],\tau_1,\tau_2)&=- \frac{b}{J^{1/2}}\,\mathrm{sgn}(\tau_1-\tau_2)\,
   \frac{e^{\frac{1}{4}(\phi(\tau_1)+\phi(\tau_2))}}
   {\Bigl(h(\tau_1) - h(\tau_2)+ \delta\Bigr)^{1/2}}=\cr
      &=- \frac{b}{\sqrt{\pi}J^{1/2}}\,\mathrm{sgn}(\tau_1-\tau_2)\,e^{\frac{1}{4}(\phi(\tau_1)+\phi(\tau_2))}
   \int_0^\infty \frac{d \alpha}{\sqrt\alpha}\,e^{-\alpha \int_{\tau_1}^{\tau_2}ds \,e^{\phi(s)} - \alpha \delta}.
\end{align}
This representation will be an important building block in all subsequent
calculations. For later reference we note that the integration over the auxiliary
variable comes with a convergence generating factor, $\sim \exp(-\alpha \delta)$,
where $\delta\sim J^{-1}$ is of the order of the inverse UV cutoff. The reason is
that we are operating within the framework of an effective low energy theory and
times of the order $\sim J^{-1}$ cannot be effectively resolved. This translates to a
smearing of the same order in the arguments of the Green functions,
$h(\tau_1)-h(\tau_2)+\delta$, which after exponentiation acts as a convergence generator.

In this form, the Green functions can now be substituted into Eq.~\eqref{eq:observable1} and we obtain the representation
\begin{align}
    \label{eq:OTOPI}
    G_4(\tau_1,\tau_2,\tau_3,\tau_4) &=\frac{b^2}{\pi J Z}
\int\limits_0^\infty  \frac{d\alpha_1 d\alpha_2}{\sqrt{ \alpha_1\alpha_2} } \!\!\int\limits_{\phi(\pm \frac \beta 2) = \phi_0} \!\!
{\cal D}\phi\, e^{\frac{1}{4}(\phi(\tau_1)+ \phi(\tau_2)+   \phi(\tau_3)+ \phi(\tau_4))} e^{-S[\phi] - S_{\alpha_1}[\phi]-S_{\alpha_2}[\phi]},
\end{align}
where
\begin{align}
    \label{eq:QuenchAction}
    S_{\alpha_1}[\phi]\equiv \alpha_1 \int\limits_{\tau_2}^{\tau_1} \!\!d\tau\, e^{\phi(\tau)},\qquad S_{\alpha_2}[\phi]\equiv \alpha_2 \int\limits_{\tau_4}^{\tau_3} \!\!d\tau\, e^{\phi(\tau)},
\end{align}
and $S[\phi]$ is given in Eq.~\eqref{eq:K_QM}. 

Eq.~\eqref{eq:OTOPI} and \eqref{eq:ActionScaled} define the path integral we need to
compute. Its  exponent  suggests an interpretation  in terms of
an effective \emph{quench dynamics}:\footnote{
Quantum quench is a protocol implying that parameters of a system are suddenly changed, thus at times $t<0$ a unitary evolution of the systems proceeds according to the Hamiltonian $\hat H_0$ while at times $t>0$ it is governed by the perturbed Hamiltonian $\hat H \neq \hat H_0$. 
In general, such sudden change of $\hat H_0$ is not supposed to be weak.
See e.g. Ref.~\cite{Calabrese:2016} for the recent review.} between the times $\tau_4$, $\tau_2$,$\tau_3$ and $\tau_1$, respectively, the strength of the Liouville potential
effectively jumps as $\gamma, \gamma+\alpha_2,\gamma+\alpha_1+\alpha_2,\gamma+\alpha_1,\gamma$ (cf. the right panel in Fig.~\ref{fig:Contour}).  
In the following sections we analyze this kink dynamics by stationary phase methods (short times/high temperatures) and via the corresponding Schr\"odinger equation (long times/low temperatures), respectively.

\section{OTO Correlation function II: short times/high temperatures} 
 \label{sec:OTOStationary}

 The computation of the OTO path integral by stationary phase methods parallels that employed in section~\ref{sec:SaddePoint} for the partition sum. 
To simplify the subsequent calculations we again set $\beta\equiv \pi$. We also re-introduce the scaled  integration variable~\eqref{eq:VarphiScaling} and in addition scale the auxiliary integration variables as
\begin{align}
    \label{eq:ScalingIntVariables}
     \alpha_i \to \alpha_i \frac{\gamma}{2M}.
\end{align}
This leaves the  integral invariant except that the action now assumes the form~\eqref{eq:ActionScaled}.

 Our later analysis will self-consistently show that the characteristic values of the
 integration variables, $\alpha_i=\mathcal{O}(1)\ll MT$. This means that 
 'quench potentials' $\alpha_i e^\varphi$ are weak as compared to the 
 unperturbed Liouville potential $e^\varphi$,
 and that it makes sense to expand around a stationary phase
  solution $\delta_\varphi S[\varphi]=M(-\varphi''+2e^\varphi)=0$ ignoring the former. The
  solution to this equation is given by $\bar\varphi=-\ln(\cos^2(\varphi))$, i.e. by
  Eq~\eqref{eq:phi_opt} neglecting the shift parameter $\lambda$ which is
  inessential except for an infinitesimal neighborhood of the integration boundaries.
  If we evaluate the functional integral on this configuration, the  integration over
  the auxiliary variables yields,
 \begin{align}
     G_4^{(0)}(\tau_1,\tau_2,\tau_3,\tau_4)&=Z^{-1}\int\limits_0^\infty  \frac{d\alpha_1 d\alpha_2}{\sqrt{ \alpha_1\alpha_2} }G(\tau_1,\tau_2|\alpha_1)G(\tau_3,\tau_4|\alpha_2)=
     G_\beta(\tau_1,\tau_2)G_\beta(\tau_3,\tau_4),
 \end{align}
 where we defined
 \begin{align}
    \label{eq:Galpha}
     G(\tau,\tau'|\alpha)\equiv -\left( \frac{b^2}{\pi J\cos\tau\cos\tau'} \right)^{1/2}e^{-\alpha(\tan\tau-\tan\tau')}.
 \end{align}
 and in the final expression encounter  the uncorrelated product of two mean-field
 Green functions~\eqref{eq:G-beta}. This suggests that fluctuations around the mean
 field will have two distinct effects: they will modify  the form of the individual
 Green functions and, more importantly, generate correlations between them. To
 disentangle these influences, we introduce the ratio
\begin{align}
    \label{eq:fRatio}
    f(\tau_1,\tau_2,\tau_3,\tau_4)\equiv \frac{G(\tau_1,\tau_2,\tau_3,\tau_4)}{G(\tau_1,\tau_2)G(\tau_3,\tau_4)},
\end{align}
where $G(\tau,\tau')$ are the mean field Green functions corrected by fluctuations. 

The leading (in $M^{-1}$) contribution to the fluctuation action comes from the quadratic expansion of the action, $S[\bar \varphi+\delta\varphi]$ in $\delta \varphi$,
\begin{align}
    \label{eq:SflOTO}
    S[\delta\varphi]=\frac{M}{2}\int_{-\pi/2}^{\pi/2} d\tau\,\delta\varphi\left(-\partial^2_\tau+\frac{2}{\cos^2\varphi}\right)\delta\varphi.
\end{align}
In addition, the fluctuation offset $\delta\varphi$ appears in the pre-exponential
factors,  and in the quench potentials $\alpha_i e^\varphi$. These multiple
appearances make the full integration over fluctuations technically cumbersome. It
turns out, however, that for large $M$ fluctuations of the pre-exponential factors
$\exp(\delta \varphi(\tau_i)/4)$ contribute only negligibly to the both the
renormalization of the Green functions and to correlations, and may be neglected. The
expansion of the quench actions $S_{\alpha_i}[\bar
\varphi+\delta\varphi]=S_{\alpha_i}[\bar
\varphi]+\sum_{n=1,2}S_{\alpha_i}^{(n)}[\bar \varphi,\delta \varphi]$, where terms with
superscript $(n)$ are of $n$th order in $\delta \varphi$, produces two contributions,
$\langle S_{\alpha_i}^{(2)}\rangle -\tfrac{1}{2} \langle
(S_{\alpha_i}^{(1)})^2\rangle$, where the angular brackets denote functional
averaging over the fluctuation action~\eqref{eq:SflOTO}. These  lead to a weak
renormalization of the $i$th Green function $G_\beta$ which drops out when the
ratio~$f$ is built. The term of interest which does generate non-vanishing
correlations is the functional average of the cross ratio
\begin{align}
    \label{eq:CrossCorrelationAction}
    \left\langle S^{(1)}_{\alpha_1}[\bar \varphi,\delta\varphi]S^{(1)}_{\alpha_2}[\bar \varphi,\delta\varphi]\right\rangle= \alpha_1\alpha_2 I(\tau_1,\tau_2,\tau_3,\tau_4)\equiv \alpha_1 \alpha_2\int_{\tau_2}^{\tau_1}d\tau \int_{\tau_4}^{\tau_3}d \tau' e^{\bar \varphi(\tau)+\bar \varphi(\tau')} \langle \delta\varphi(\tau)\delta \varphi(\tau')\rangle.
\end{align}
The straightforward but lengthy computation of the functional expectation value is detailed in~\ref{sec:CrossCorrelation} and yields
\begin{align}
    \label{eq:IIntegralResult}
     I(\tau_2,\tau_4)\simeq -\frac{\sin(4\tau_2)\cos(2\tau_4)\cos^2\tau_4}{2 M\sin^2(2\tau_2)\sin^2(2\tau_4)}+(\tau_2\leftrightarrow\tau_4),
 \end{align} 
 where we used that (including during the analytic continuation) we may impose the constraint $\tau_3=\tau_4+\pi/2$ and
$\tau_1=\tau_2+\pi/2$, and the $\simeq$ sign indicates the omission of terms exponentially vanishing under under the continuation to large imaginary times $it \gg i\beta$. Using this result, we find that the leading contribution in $M^{-1}$ to the correlation ratio Eq.~\eqref{eq:fRatio} is given by
\begin{align}
    f(\tau_2,\tau_4)=1+\frac{I(\tau_2,\tau_4)}
   { [G_\beta(\pi/2)]^2 }
\int\limits_0^\infty  \frac{d\alpha_1 d\alpha_2}{\sqrt{ \alpha_1\alpha_2} } \alpha_1 \alpha_2 G(\tau_1,\tau_2|\alpha_1)G(\tau_3,\tau_4|\alpha_2).
\end{align}
Substituting Eq.~\eqref{eq:Galpha}, the $\alpha$-integrals appearing in this expression become
\begin{align}
    \frac{1}{G_\beta(\tau_1,\tau_2)}\int\limits_0^\infty  \frac{d\alpha_1}{\sqrt{ \alpha_1} }\, \alpha_1  G(\tau_1,\tau_2|\alpha_1)=\left.\frac{1}{2(\tan\tau_1-\tan\tau_2)}\right|_{\tau_1=\tau_2+\pi/2}=\frac{\sin2\tau_2}{4},
\end{align}
and analogously for the second integral. We substitute them in Eq.~\eqref{eq:IIntegralResult} to obtain
\begin{align}
    f(\tau_2,\tau_4)&=1-\frac{\left( \sin(4\tau_2)\cos(2\tau_4)\cos^2\tau_4+(\tau_2\leftrightarrow\tau_4) \right)}{32 M\sin(2\tau_2)\sin(2\tau_4)}\longrightarrow\cr
    &\longrightarrow 1-\frac{e^{2t}}{64 M}+\mathcal{O}(e^{t}/M),
\end{align}
where in the  final step we have performed the analytic continuation to the time
arguments (cf. Eq.~\eqref{eq:C4Times}) $\tau_{2}\to -\pi/4-it/2$,  $\tau_4\to - \pi/2 +it/2$. 
Expressed in the original language of correlation functions, and re-introducing the temperature parameter, this result assumes the form
\begin{align}
    \label{eq:FShortTime}
    F(t)=G_{\beta}(\beta/2)^2 \left( 1-\frac{\beta e^{2\pi t/\beta}}{64 \pi M}+\mathcal{O}(e^{\pi t/\beta }/M) \right).
\end{align}
This agrees, including pre-factors, with Eq.~(6.59) of Ref.~\cite{Maldacena1:2016},  where it was obtained from the Schwarzian action, without recourse to its Liouvillian reformulation. This expression may be trusted up until the second term in the brackets becomes comparable with unity, which happens at the Ehrenfest time $t_E$, given by Eq.~(\ref{eq:Ehrenfest}).

\section{OTO correlation function III~\ref{sec:SaddePoint}: intermediate times/high temperatures}
\label{sec:OTOIntermediateTimes}

In this section we briefly discuss what happens for times larger than the Ehrenfest
time $t_E$, Eq.~(\ref{eq:Ehrenfest}), at which the correction discussed previously overpowers the leading
order result (\ref{eq:FShortTime}). Yet we restrict ourselves to times smaller than $\sim M$, where the strong Goldstone mode
fluctuation regime is entered. This intermediate regime is still amenable to
stationary phase methods. However, the technicalities get a little complicated because the quench 
potentials $\alpha_i e^\varphi$ can no longer be neglected in the
solution of the stationary phase equations. Referring
to~\ref{sec:StationaryPhaseIntermediate} for details, the stationary phase procedure
leads to the intermediate result for OTO correlation with fixed $\alpha_{1,2}$: 

\begin{align}
    \label{eq:FIntermediateIntermediate}
    F(t|\alpha_1,\alpha_2)&=  \frac{8M b^2\,e^{-\frac{i\pi}{4}}}{\pi J}  \frac{\omega^2 \,e^{-\omega t+2 \pi M (\omega^2-1)}}{((1+\alpha_1)(1+\alpha_2)(1+\alpha_1+\alpha_2))^{1/4}},\crcr
    &\qquad \qquad \omega=1-\frac{i}{\pi}\ln \frac{1+\alpha_1+\alpha_2}{(1+\alpha_1)(1+\alpha_2)}.
\end{align}
This formula already contains the essence of the result: exponential decay of the correlation function with an exponent whose real part scales as $\sim \exp(-t)\to \exp(-t \pi/\beta)$. However, if we aim to fix the  result including pre-exponential factors, the integration over auxiliary variables 
\begin{align}
    F(t)=\int_0^\infty \frac{d\alpha_1 d\alpha_2}{\sqrt{\alpha_1\alpha_2}}e^{-\alpha_1-\alpha_2}F(t|\alpha_1,\alpha_2).
\end{align}
 needs to be performed. Here, we recalled that the $\alpha$-integrations come with a
 convergence generating factor $\delta\sim J^{-1}$ (cf. discussion after
 Eq.~\eqref{eq:GreenFunctionAlpha}) which under the rescaling by $\gamma\sim J$ gets
 promoted to a factor of $\mathcal{O}(1)$. This factor acts to regularize an
 otherwise logarithmically divergent integral. The integration is not entirely
 trivial. One observes that the integrand does not couple to physical parameters,
 except through the variable combination $\omega(\alpha_1,\alpha_2)$. This suggests
 to introduce a new variable, $\lambda=\lambda(\alpha_1,\alpha_2)\equiv \ln
 ((1+\alpha_1)(1+\alpha_2)/(1+\alpha_1+\alpha_2))$. Expressed in terms of $\lambda$,
 the exponent becomes Gaussian, and the rest of the integral, $\int d\alpha_1
 d\alpha_2 (\dots) \delta(\lambda(\alpha_1,\alpha_2)-\lambda)\equiv \rho(\lambda)$
 turns into an effective `spectral density'. Using that for large $M$ only small
 deviations of $\lambda$ off zero are relevant, one may verify that
 $\rho(\lambda)\simeq \ln(\lambda^{-1/2})/\sqrt{\lambda}$, and the integral becomes
 \begin{align}
      F(t)\sim \frac{M}{\pi J}e^{-t - i \pi/4}\int_0^\infty d\lambda\, \rho(\lambda)e^{4 i M \lambda}\sim \frac{M}{\pi J} \sqrt{M}\ln(M) e^{-t}.
  \end{align}
  Finally, re-introducing $\beta$ and recalling the value of the finite temperature Green functions $G_\beta(\beta/2)\sim (\beta J)^{-1/2}$, we obtain
  \begin{align}
       \label{eq:FIntermediateFinite}
       F(t)\sim G^2_\beta(\beta/2) \ln \left( \frac{M}{\beta} \right)\, e^{-\pi (t-t_E)/\beta},
   \end{align} 
where the definition of the Ehrenfest time, $\exp(\pi t_E/\beta) = \sqrt{M/\beta}$ was used. To exponential accuracy this agrees with the black hole shock wave analysis of Ref.~\cite{Maldacena1:2016} which predicted a decay $\sim t\exp(-\pi t/\beta)$.

For completeness we mention that the above discussion assumed a fixation of the real
parts $\mathrm{Re}(\tau_i)$ of the time arguments relative to that,
$\tau^\ast=\beta/2$, of the singularity which is necessarily present in a bijective
transformation $[-\beta/2,\beta/2]\to \Bbb{R}$ (cf. discussion in the beginning of
section~\ref{sec:SoftModeIntegration}.) This fixation may be abandoned by introducing
a shift parameter $\tau_i\to \tau_i +s$ and integrating over the latter (cf.
Fig.~\ref{fig:Contour}.) On top of that, one may place the singularity in-between any
of the arguments $\tau_i$, and sum over all four choices. While the relative ordering of the singularity and the observation times turns out to be irrelevant, the continuous shift parameter begins to play a role when we
turn to the complementary regime of large observation times:

\section{OTO correlation function IV: long times/low temperatures} 
\label{sec:OTOLongTimes}

The previous sections served benchmarking purposes where we reproduced known results within the framework of the finite temperature extension of Liouville theory. We now turn to uncharted territory and address the theory at low temperatures, where quantum fluctuations are large and different methodology is required. As in section~\ref{sec:PartitionQM}, the strategy is to turn a foe into a friend and use that largeness of the $\phi$-fluctuations implies confinement of the fluctuations of its conjugate momentum, $k$. 

Once more the calculation simplifies if a convenient set of complex time arguments is chosen. Presently, we find it advantageous to start from the configuration~\eqref{eq:C4Times} at $it=\tau\in \Bbb{R}$.
Here, $\tau$ serves as a real time parameter which will eventually be continued into the complex plane, and $s\in [0,\beta/4]$ is the parameter controlling the relative position of the observation times and the time, $\tau^\ast$, of the reparameterization singularity (cf. discussion at the end of the previous section). 
The correlation function is then obtained as
\begin{align}
    F(t)\equiv \mathrm{Re}(F(\tau)\big|_{\tau\to i t})\equiv \mathrm{Re}\frac{1}{\beta}\int_0^{\beta/4+\tau}ds \, G_4(\tau_1,\tau_2,\tau_3,\tau_4)\big|_{\tau\to i t}+\dots,
\end{align}
where $G_4$ is given by Eq.~\eqref{eq:OTOPI}, and the integral implements an average
over the shift parameter, $s$. The ellipses stand for the three contribution of
different ordering of the singularity relative to the observation times. One can explicitly verify that all four
orderings give identical result for the real part while a pairwise cancellation of imaginary parts occurs. 
Therefore we will focus on the one explicitly
displayed throughout an account for the remaining ones by a multiplicative factor of $4$. 

It will be convenient to scale the integration variable in the path integral as
$\phi\to \phi+\ln \gamma$, and the auxiliary variables as $\alpha_i\to
\alpha_i/\gamma$. This operation leaves the form of the path integral invariant but
changes the strength of the potentials in the Liouville action as $\gamma\to 1$, and
$\alpha_i\to \alpha_i/\gamma$. As in section~\ref{sec:PartitionQM}, we re-interpret the path integral in a
Schr\"odinger picture where it describes time evolution under the Hamiltonian
Eq.~\eqref{eq:Hamiltonian_QM}. However, the presence of the quench potentials 
$\alpha_{1,2} e^\varphi$ implies that the strength of the Liouville potential jumps from $1$ to
$1+\alpha_1/\gamma$ and $1+\alpha_2/\gamma$ in the time intervals $[\tau_2,\tau_1]$
and $[\tau_4,\tau_3]$, respectively (cf. Eq.~\eqref{eq:QuenchAction}.) This requires
the insertion of different sets of eigenfunctions labeled as
$|k\rangle,|k^{(\alpha)}\rangle$, depending on whether the quench potentials are
absent or present with value  $\alpha$, respectively.

Specifically, the presence of four observation times and one singularity requires the insertion of five resolutions of unity. The spectral representation of the correlation function then assumes the form
\begin{align}
    F(\tau)&=\frac{b^2}{\pi J Z(\beta)}\frac{4}{\beta}\int_0^{\beta/4+\tau}ds \int_0^\infty \frac{d\alpha_1d\alpha_2}{\sqrt{\alpha_1\alpha_2}}\sum_k e^{-E_{k_1}(\tau_4+\frac{\beta}{2})-E_{k_2}\tau_{24}-E_{k_3}\tau_{32}-E_{k_4}\tau_{13}-E_{k_5}(\frac{\beta}{2}-\tau_1)}\times\crcr
    &\qquad \times \langle \phi_0|k_1\rangle\,\langle k_1|e^{\frac{\phi}{4}}|k^{(\alpha_2)}_2\rangle\, \langle k^{(\alpha_2)}_2|e^{\frac{\phi}{4}}|k^{(\alpha_1+\alpha_2)}_3\rangle\, \langle k^{(\alpha_1+\alpha_2)}_3|e^{\frac{\phi}{4}}|k^{(\alpha_1)}_4\rangle\, \langle k^{(\alpha_1)}_4|e^{\frac{\phi}{4}}|k_5\rangle\,\langle k_5|\phi_0\rangle,
\end{align}
where $\tau_{ij}=\tau_i-\tau_j$ are the time differences 
\begin{align}
    \label{eq:TauDifferences}
    \tau_{24} = \tau_{13} = \frac \beta 4 - \tau, \qquad \tau_{32} = \frac \beta 4  + \tau.
\end{align}
Central to the expression above are the matrix elements, $\langle k^{(\alpha)}|e^{\frac{\phi}{4}}|k^{'(\alpha')}\rangle$ between eigenstates of different Liouville potential. In ~\ref{sec:ComputationMatrixElements} we show that for the small momenta $k$ relevant to the correlation function at large times they show a high degree of universality. The matrix elements factorize into products of the $\alpha$-independent normalization factors ${\cal N}_k= \tfrac{2}{\Gamma( 2 i k)}$ (cf. Eq.~\eqref{eq:eigenfunction}) and $M^{-1/4}$ times a term which depends \emph{only} on $\alpha$. The integral over these variables can thus be performed to yield a numerical factor $C=\mathcal{O}(1)$. This leads to the intermediate result  
\begin{align}
    F(\tau)&=\frac{C b^2}{\pi M J Z(\beta)}\frac{4}{\beta}\int_0^{\beta/4+\tau}ds \prod_{i=1}^5 \int \frac{dk_i}{2\pi}|\mathcal{N}(k_i)|^2 e^{-E_{k_1}(\tau_4+\frac{\beta}{2})-E_{k_2}\tau_{24}-E_{k_3}\tau_{32}-E_{k_4}\tau_{13}-E_{k_5}(\frac{\beta}{2}-\tau_1)}.
\end{align}
Here, we used that for the diverging boundary conditions, $\phi_0\to \infty$, the
boundary wave functions, $\langle
\phi_0|k\rangle=\mathcal{N}(k)K_{2ik}(2\sqrt{2M}e^{\phi_0})\to \mathcal{N}(k)\times
\mathrm{const.}$, converge to a product of the normalization factor and a
$k$-independent term. The latter  cancels against the corresponding factor appearing
in the partition sum (see below) and is ignored.

Using the  expansion $|\mathcal{N}(k)|^2=|\Gamma(2ik)|^{-2}\simeq 4k^2$, and the auxiliary relation 
$\int \frac{dk}{2\pi} \, k^2 \exp(-E_k \tau)=\tfrac{1}{\sqrt{2\pi}}(M/\tau)^{3/2}$, we obtain the proportionality,
\begin{align}
     F(\tau)&\sim \frac{M^{13/2}}{Z(\beta)J}\frac{1}{(\tau_{13}\tau_{32}\tau_{24})^{3/2}}\frac{4}{\beta}\int_M^{\beta/4+\tau-M}ds\frac{1}{s^{3/2} \left( \frac{\beta}{4}-s+\tau \right)^{3/2}}+\dots\sim\crcr
     &\sim \frac{M^{6}}{Z(\beta)J}\frac{1}{(\tau_{13}\tau_{32}\tau_{24})^{3/2}}\frac{1}{\beta} \frac{1}{\left(\frac{\beta}{4}+\tau \right)^{3/2}}.
     \label{eq:Ft_prop}
 \end{align} 
 Here, the ellipses $\dots \sim\int_0^Mds(\dots)$ refer to the contribution to the integral where the observation points are close to the boundary. The short time distance Green functions appearing in this regime can no longer be described within the low momentum approximation used above. However, referring to Ref.~\cite{Bagrets:2016} for details we note that they behave  as $\sim 1/Ms^{1/2}$, matching the above $s^{-3/2}$ law at $s\sim M$. 
In this way the  boundary contributions effectively regularize the $s^{-3/2}$ singularity of the integral.
We  also note that the accumulated appearance of $\tau^{-3/2}$ scaling in this
relation is a hallmark of Liouville quantum mechanics. As mentioned in the
Introduction, cf. Eq.~(\ref{eq:OTOLehmann}), it may be qualitatively understood assuming statistical independence of many-body energies and matrix elements and utilizing square root singularity in the many-body DoS. 

Finally, noting the scaling of the partition sum $Z(\beta)\sim
\int dk |\langle \phi_0|k\rangle|^2\exp(-\beta k^2/2M)\sim (M/\beta)^{3/2}$,
substituting the time arguments, Eq.~\eqref{eq:TauDifferences}, and continuing to
real times, $\tau\to it$, we obtain
\begin{align}
    F(t)\sim \frac{M^{9/2}\beta^{1/2}}{J}\frac{1}{\left( \left(\frac{\beta}{4}\right)^2+t^2 \right)^{3}}
    \stackrel{t\gg \beta}\sim t^{-6}.
    \label{eq:Ft_long_t_low_T}
\end{align}
Incidentally, the $t^{-6}$ power law scaling coincides with that found for
non-interacting spinless fermions \cite{Dora:2016}. 

The application of the same spectral decomposition procedure to the   Greens function, $G(\tau)$, yields the result 
\begin{equation}
\label{eq:QM_G2_4}
G(\tau)  \sim  - \,\frac  {M^2 \beta^{1/2}} {\sqrt{J}} \, \frac{{\rm sgn}(\tau)}{\tau^{3/2}(\beta-\tau)^{3/2}}, \qquad \tau \gg M.
\end{equation}
Using this formula to  normalize the OTO function, $F(t) \to F(t)/[G(\beta/2)]^2$, we obtain the result describing  regime 4 in Eq.~\eqref{eq:Summary}.

The results derived above can readily be extended to the regime of  long times but high temperatures, $ \beta \ll M \ll t$.
To this end we substitute into Eq.~(\ref{eq:Ft_prop}) the high temperature expression for the partition sum~(\ref{eq:Z_sem}), and use the scaling 
 $G(\beta/2) \sim (J\beta)^{-1/2}$ of the 2-point Greens function to  normalize.  As a result, Eq.~(\ref{eq:Ft_long_t_low_T}) changes to
\begin{equation}
F(t)/[G(\beta/2)]^2 \sim e^{-2\pi^2 M/\beta} \left(\frac \beta M \right)^{3/2} \left(\frac M  t \right)^{6} \propto t^{-6}, \qquad \beta \ll M \ll t,
\end{equation}
which now describes the OTO function in the regime 3 of Eq.~(\ref{eq:Summary}).

\section{Discussion}
\label{sec:Discussion}

In this paper we explored the role played by large conformal Goldstone mode
fluctuations of the SYK model within a finite temperature framework suitable for the
study of out of time order correlations. Our theory is constructed by exploiting the
extensive time-reparameterization freedom of the SYK model. This allows to describe
Goldstone mode fluctuations in terms of a real valued field defined on the periodic
imaginary time interval which, however, necessarily contains a point of singularity.
The handling of this singularity required some care and we tested the theoretical
framework by comparison to various known, or conjectured results, including the
temperature dependence of the many-body partition sum, the exponential buildup of OTO
de-correlations at short times, and their exponential decay at times larger than the
Ehrenfest time. 

The truly novel content of the theory unfolds at times larger than a
scale $M\sim N\ln(N) J^{-1}$, where $N$ is the number of particles in the system or,
equivalently, a logarithmic measure for the dimension $\sim 2^{N/2}$ of the many-body Hilbert
space. This regime is inaccessible to theoretical approaches taking thermodynamic
limits $N\to \infty$ at a fixed time, but arguably may play a key role in the
description of the regularizing effects of quantum mechanics (via finite $N$) on long
time fluctuations. Our main finding is that in such regimes, and equally in regimes
of low temperatures, $T<M^{-1}$ at generic times $t>T^{-1}$, four-point OTO correlations decay with the
universal power law $\sim t^{-6}$. 

While the  calculations required to nail this behavior are  technical in parts (mainly due to the required careful handling of  the singularity), the  $6=4\times 3/2$ power law has its  origin in
the celebrated $t^{-3/2}$ universality of the Liouville quantum mechanics \cite{Shelton:1998}. It is a robust
feature of this system that temporal correlations of local operators decay as $\langle \hat O(0)\hat
O(t)\rangle\sim t^{-3/2}$. This power-law may also be interpreted in terms of the $\sqrt{E}$ scaling of the average low-energy many-body DoS, Eq.~\eqref{eq:DoS}, as discussed in the Introduction.  
The quadrupling of the $3/2$ exponent has to do with the definition of the 
OTO correlation function whose particular time ordering of operators requires the
introduction of four temporal contours, Fig.~\ref{fig:Contour}. This time arrangement makes all {\em four} time arguments of the correlation function (\ref{eq:Green-4}), (\ref{eq:C4Times}) to be separated by long intervals $\sim |t\pm i\beta/4|$. 
One may generalize this observation and claim that any multipoint OTO correlation is given by the product of $3/2$ powers of all 
long  ($\gg M$)  time intervals involved. 

\vskip.3cm
\noindent  \emph{Note added:} 
Recent preprint~\cite{Stanford:2017} discusses one-loop (semiclassical) exactness of the partition sum of the Schwarzian theory~(\ref{eq:Z_sem}) using
the Duistermaat-Heckman formula~\cite{Duistermaat:1982}.  It remains to be seen if this technique is also helpful in evaluation of correlations functions.

\vskip.3cm
\noindent  \emph{Acknowledgments:} We thank J. Verbaarschot and A.M. Garc\'ia-Garc\'ia for discussions and for sharing unpublished numerical data. A.K. was supported by DOE contract DEFG02- 08ER46482, and acknowledges support by QM2--Quantum Matter and Materials of Cologne University. Work supported by CRC 183 of the Deutsche Forschungsgemeinschaft (project A03).

\appendix

\section{Derivation of Eq.~\eqref{eq:action}}
\label{sec:StartingAction}

In this appendix, we briefly review how the action~\eqref{eq:action} emerges as an effective description of the system described by the Hamiltonian~\eqref{eq:Hamiltonian}. To this end, consider a Grassmann coherent state functional integral representation of the replicated partition sum $Z^n\equiv (\mathrm{tr}(\exp(-\beta \hat H))^n$,
\begin{align*}
    Z^n=\int D(\chi)\,e^{-\sum_a^n\int_0^\beta d\tau\, (\chi_i^a\partial_\tau \chi_i^a+\hat H(\chi^a))},
\end{align*}
where $\chi^a_i$ are $Nn$ independent Grassmann variables, $a=1,\dots,n$ is a replica index, and the replication is introduced to get rid of the partition sum after observables have been computed, $\lim_{n\to 0}Z^n=1$. Averaging over disorder, we obtain 
\begin{align}
    \left\langle Z^n\right\rangle&=\int D(\chi)\,e^{-\sum\limits^n_{a}\int_0^\beta d\tau\, \chi_i^a\partial_\tau \chi_i^a+\frac{J^2}{8N^3} \sum\limits^n_{a,b} \int d\tau  d\tau' \sum\limits^N_{ijkl}
\chi_i^a(\tau)\chi_j^a(\tau)\chi_k^a(\tau)\chi_l^a(\tau)\times \chi_i^b(\tau')\chi_j^b(\tau')\chi_k^b(\tau')\chi_l^b(\tau')}=\cr
&=\int D(\chi)\,e^{-\sum\limits^n_{a}\int_0^\beta d\tau\, \chi_i^a\partial_\tau \chi_i^a+ N \frac{J^2}{8} \sum\limits^n_{a,b} \int d\tau d\tau' \big[\tilde G^{ab}_{\tau,\tau'}\big]^4},
\end{align}
where 
$\tilde G^{ab}_{\tau,\tau'}=- N^{-1}\sum\limits_i^N \chi_i^a(\tau)\chi_i^b(\tau')$.
To pass  from the abbreviation, $\tilde G$, for a Grassmann bilinear to an integration variable, $G$, we introduce 
\begin{equation}
1=\int \mathcal{D}G\,\, \delta\Big( N G^{ab}_{\tau,\tau'} + \sum\limits_i^N \chi_i^a(\tau)\chi_i^b(\tau')\Big) =
\int \mathcal{D}G\, \mathcal{D} \Sigma\,\, e^{ {1\over 2}\sum\limits^n_{a,b} \int d\tau d\tau'    
\Sigma^{ba}_{\tau',\tau}
\Big(N G^{ab}_{\tau,\tau'} + \sum\limits_i^N \chi_i^a(\tau)\chi_i^b(\tau')\Big)}
\end{equation}
in the functional integral, which entitles us to replace $\tilde G\to G$ in the quartic term. As a result, the action has become quadratic in Grassmann variables, and the Gaussian integration over these produces the Pfaffian $\left|\partial_\tau \delta^{ab} + \Sigma^{ab}_{\tau,\tau'}\right|^{N\over 2}$. We re-exponentiate the latter to obtain the representation
\begin{align*}
    \left\langle Z^n\right\rangle=\int \mathcal{D}G\, \mathcal{D} \Sigma\, e^{-S[\Sigma,G]},
\end{align*}
where the effective action is given by Eq.~\eqref{eq:action}.

\section{Remarks on the Lagrange multiplier}
\label{sec:LagrangeRemarks}

The job of the Lagrange multiplier is to establish, in the best possible way, the
constraint~\eqref{eq:t-H}. It can be understood in analogy $\gamma\leftrightarrow
\mu$ to the chemical potential fixing the particle number $\int e^\phi\leftrightarrow
N$ to a premeditated value $t_H \leftrightarrow N_0$ in statistical mechanics.
Furthering this analogy, fluctuations away from the intended value $\Delta
t_H[\phi]\equiv \int e^\phi-t_H$ are obtained as expectation values $\langle \Delta
t_H[\phi]\rangle_\phi= - \partial_\gamma \ln Z(\gamma)$, where $Z(\gamma)\leftrightarrow
Z(\mu)$ is the functional partition sum depending on the chosen value of $\gamma$,
and fluctuations as
\begin{align}
     \mathrm{var}(\Delta t_H[\phi])=\partial_\gamma^2 \ln Z(\gamma). 
 \end{align} 
The evaluation of these formulas on the partition sum~\eqref{eq:Z_sem_full}
yields $\langle \Delta t_H[\phi]\rangle_\phi=\gamma^{-1}$ and $\mathrm{var}(\Delta
t_H[\phi])=\gamma^{-2}$, respectively. This shows that large values of $\gamma$
increase the accuracy of the formalism. Our logics is to implement the constraint
through  freely adjustable parameters (notably the values $\phi_0$ of the field at
the boundaries of the temporal domain) and choose for $\gamma$ the largest value
consistent with the low energy nature of the theory, $\gamma\sim J$. For these
values, fluctuations in the constraint variables, which have dimensionality
$[\mathrm{time}]$ are of $\mathcal{O}(J^{-1})$ which is the smallest time scale
resolvable in the theory.

\section{Gaussian fluctuations}
\label{sec:Gaussian}

The straightforward Gaussian expansion of the action~\eqref{eq:ActionScaled} in $\bar\varphi+\delta \varphi$ leads to
 \begin{equation}
S_G[\delta \phi] = \frac{M}{2} \int\limits_{-\pi/2}^{\pi/2} d\tau\,\, \delta\varphi(\tau) \left[ - \partial_\tau^2 + \frac{2 (1- \lambda)^2}{\cos^2 \left( \tau(1-\lambda)\right)} \right] \delta\varphi(\tau).
\end{equation}
The resulting  fluctuation determinant is obtained by  standard
recipes~\cite{Kleinert:2009}:  consider \emph{any} two independent solutions $z_{1,2}$ of the
differential equation
\begin{equation}
\left[ - \partial_\tau^2 + \frac{2(1- \lambda)^2}{\cos^2 \left(\tau(1-\lambda)\right)} \right] z_{1,2}(\tau) = 0.
\label{eq:Diff_z}
\end{equation}
For later convenience, we choose
\begin{equation}
\label{eq:Zfunctions}
z_1(\tau) = 1 + \left( \frac \pi 2  +\tau(1-\lambda) \right) \tan(\tau(1-\lambda)), \qquad
z_2(\tau) = 1 - \left( \frac \pi 2  - \tau(1-\lambda)\right) \tan (\tau(1-\lambda)).
\end{equation}
The fluctuation determinant is then given by~\cite{Kleinert:2009}
\begin{equation}
D = \frac{1}{M\,W} \Bigl[  z_1(\pi/2) z_2(-\pi/2) - z_1(-\pi/2) z_2(\pi/2)  \Bigr] \simeq 
 \frac{4}{M \lambda^2}. 
\end{equation}
where $W= z_1' z_2  - z_1 z_2' = \pi(1-\lambda)$ is the time-independent Wronskian and $M$ features as the effective inverse `Planck-constant'~\cite{Kleinert:2009}.
Multiplying  the exponentiated saddle point action with the factor $D^{-1/2}$ 
and re-introducing  $\beta$  we obtain  
\begin{equation}
\label{eq:Z_sem_full}
Z(\beta) \sim  \frac{1}{\gamma t_H} \left(\frac{M}{\beta} \right)^{3/2}\exp\left( \frac{2 \pi^2 M}{\beta}\right)
\end{equation}
which is Eq.~\eqref{eq:Z_sem} up to the normalization prefactor.

\section{Computation of the cross correlation Eq.~\eqref{eq:CrossCorrelationAction}} 
\label{sec:CrossCorrelation}

The central object in Eq.~\eqref{eq:CrossCorrelationAction} is the expectation value of the fluctuation field,
which assumes the form
\begin{align}
    \label{eq:FluctuationCorrelator}
    \left\langle \delta\varphi(\tau)\delta\varphi(\tau')\right\rangle\equiv P(\tau,\tau')=\frac{1}{M W}\left( z_1(\tau)z_2(\tau')\Theta(\tau'-\tau)+ z_2(\tau)z_1(\tau')\Theta(\tau-\tau')\right),
\end{align}
where $z_{1,2}$ are independent solutions of the differential Eq.~(\ref{eq:Diff_z}) satisfying the boundary conditions
$z_1(-\pi/2) = 0$, $z_2(\pi/2) = 0$  and $W$ is their Wronskian. For times $\tau$ in the bulk of the interval, the boundary regulators $\lambda$ can be neglected and the functions $z$'s entering the fluctuation correlator become 
\begin{align}
 z_1(\tau) = 1 + \left( \frac \pi 2  + \tau \right) \tan  \tau, \qquad
z_2(\tau) = 1 - \left( \frac \pi 2  -  \tau \right) \tan \tau,
\end{align}
with $W=\pi$. These expressions need to be substituted into Eq.~\eqref{eq:CrossCorrelationAction} and integrated as 
\begin{align}
    I(\tau_1,\tau_2,\tau_3,\tau_4)\equiv\int_{\tau_2}^{\tau_1}d\tau \int_{\tau_4}^{\tau_3}d \tau' \frac{P(\tau,\tau')}{\cos^2(\tau)\cos^2(\tau')} ,
    \label{eq:I_4t}
\end{align}
where $\bar\varphi=\ln(\cos^{-2}\tau)$ was used. 
Although this is straightforward in principle,  the
evaluation is cumbersome in practice. Calculation becomes simpler if we note that (cf.
Eq.~\ref{eq:C4Times}) we may set $\tau_3=\tau_4+\pi/2$ and
$\tau_1=\tau_2+\pi/2$.  
Prior to  analytic continuation the time arguments lie on the real axis and are ordered as
 $\tau_4<\tau_2<\tau_3<\tau_1$ according to the values of the readout
times~\eqref{eq:C4Times}. This structure suggests to define  integrals over simplified `quadratic' integration domains, 
\begin{align}
    J(\tau_a,\tau_b)\equiv \frac 1 2 \int_{\tau_a}^{\tau_b} d\tau \int_{\tau_a}^{\tau_b} d \tau'\frac{P(\tau,\tau')}{\cos^2(\tau)\cos^2(\tau')}=\frac{1}{\pi M}\int_{\tau_a}^{\tau_b} d\tau  \frac{z_2(\tau)}{\cos^2(\tau)} 
\int_{\tau_a}^\tau d\tau' \frac{z_1(\tau')}{\cos^2(\tau')},
\end{align}
and apply elementary geometric reasoning to show that Eq.~(\ref{eq:I_4t}) reduces to
\begin{align}
    I(\tau_2,\tau_4)=\Bigl(J(\tau_4,\tau_1)+J(\tau_2,\tau_3)-J(\tau_4,\tau_2)-J(\tau_3,\tau_1)\Bigr)\big|_{\tau_1=\tau_2+\pi/2\atop \tau_3=\tau_4+\pi/2}.
\end{align}
The situation simplifies further if we observe that eventually an analytic continuation to large imaginary values $it$ will be carried out. This implies that in the integration over trigonometric functions only those contributions need to be kept that will not vanish exponentially. Under this condition the straightforward but tedious calculation of the auxiliary integral simplifies to the result~\eqref{eq:IIntegralResult}.

\section{Computation of the matrix elements $\langle k^{(\alpha)}|e^{\frac{\phi}{4}}|k^{'(\alpha')}\rangle$} 
\label{sec:ComputationMatrixElements}

After the scaling, $\gamma\to 1$, $\alpha\to \alpha/\gamma$ the Liouville eigenfunctions in the $\phi$-representation are given by Eq.~\eqref{eq:eigenfunction} with the replacement $\gamma\to 1 + \alpha$. As a consequence, these matrix elements of the operator $\exp(\phi/4)$ assume the form
\begin{align}
 \label{eq:MatrixElementsExpOps}
W(k,k'|\alpha,\alpha')&\equiv \langle k^{(\alpha)}|e^{\frac{\phi}{4}}|k^{'(\alpha')}\rangle=\crcr
&= \overline{\mathcal{N}_k}\mathcal{N}_{k'}\int d\phi\, K_{2ik}\left(2\sqrt{2 M(1+\alpha)} e^{\phi/2}\right)\, e^{\phi/4}\,K_{2ik'}\left(2\sqrt{2 M(1+\alpha')} e^{\phi/2}\right)=\crcr
&=\frac{\overline{\mathcal{N}_k}\mathcal{N}_{k'}}{(M/2)^{1/4}}\int \frac{dz}{\sqrt{z}}\,  K_{2ik}\left(z\sqrt{1+\alpha}\right) \,K_{2ik'}\left(z\sqrt{1+\alpha'}\right).
\end{align}
For generic values of $k,k'$ the integral evaluates to an complicated configuration of hypergeometric functions. However, at large times only small momenta matter, and to zeroth order in $k$ the result simplifies to
\begin{align}
    W(k,k'|\alpha,\alpha')\simeq \frac{\overline{\mathcal{N}_k}\mathcal{N}_{k'}}{M^{1/4}}W(\alpha,\alpha'),\qquad W(\alpha,\alpha')=\frac{\Gamma^2(\tfrac{1}{4})}{2^{5/4}}\frac{1}{(1+\alpha)^{1/2}}\sum_{\pm}
K \left(\tfrac{1}{2} \pm \tfrac 1 2 \sqrt{\frac{\alpha-\alpha'}{1+\alpha}} \right) , 
\end{align}
where $K$ is the complete elliptic integral of the first kind. 
For small values of $\alpha$ the functions $K$ are regular, and for large arguments weakly decay as $\alpha^{-1/4}$. This translates to a logarithmic singularity of the $\alpha$-double integral. What comes to rescue is the convergence generating factor (cf. discussion below Eq.~\eqref{eq:GreenFunctionAlpha}). In the scaled framework, $\gamma\to 1$, this factor is upgraded as $\delta\sim J^{-1}$ to $\delta \gamma\sim 1$ and to the integral we need to perform reads  
\begin{align}
    \int_0^\infty \frac{d\alpha_1d\alpha_2}{\sqrt{\alpha_1\alpha_2}} W(0,\alpha_2)W(\alpha_2,\alpha_1+\alpha_2)W(\alpha_1+\alpha_2,\alpha_1)W(\alpha_1,0)e^{-\alpha_1-\alpha_2}\equiv C,
\end{align}
where $C$ is a constant of order unity. 

\section{Stationary phase at intermediate times} 
\label{sec:StationaryPhaseIntermediate}

At intermediate time scales the quench potentials can no longer be neglected and we
need to expand around the extrema of the action
$S[\varphi]+S_{\alpha_1}[\varphi]+S_{\alpha_2}[\varphi]$, where $S[\varphi]$ is given
by Eq.~\eqref{eq:ActionScaled} and $S_{\alpha_i}[\varphi]$ by
Eqs.~\eqref{eq:QuenchAction} with the replacement $\phi\to \varphi$. Presently, we find it convenient to scale the auxiliary variables as $\alpha_i\to 2M \alpha_i$. The entire action then is multiplied by a factor $M$, and that the functional integral picks up the same multiplicative factor. The scaling of variables also implies that the functional integral for the four point function picks up global multiplicative factor $2M$.
Due to the piecewise
constancy of the potential strength, the solutions to the stationary
phase equations $\bar\varphi''-2\gamma_i \exp(\bar\varphi)=0$, $ \gamma_0=1,\gamma_1=
1+\alpha_1,\gamma_2=1+\alpha_1+\alpha_2,\gamma_3=1+\alpha_2,\gamma_4=1$ for the five
temporal regimes defined by the time arguments
$-\beta/2,\tau_4,\tau_2,\tau_3,\tau_1,\beta/2$ assume the form $\bar
\varphi=\ln(\omega^2\gamma_i^{-1}\cos^{-2}(\omega\tau+\delta))$, where $\omega$
and $\delta$ are free parameters. Requiring the usual boundary singularity $\bar\phi(\pm \pi/2)\to +\infty$ and continuity of the solution in the bulk a straightforward fixation procedure yields for these parameters
\begin{align}
    & \qquad \delta_0 = -\delta_4 =  \frac{i}{2}\ln\frac{1+\alpha_1+\alpha_2}{(1+\alpha_1) (1+\alpha_2)},\crcr
    & 
\delta_1 = \frac{i}{2}\ln\frac{1+\alpha_1+\alpha_2}{1+\alpha_1}, \quad
\delta_2 = \frac{i}{2}\ln\frac{1+\alpha_2}{1+\alpha_1}, \quad
\delta_3 = \frac{i}{2}\ln\frac{1+\alpha_2}{1+\alpha_1+\alpha_2},
 \end{align} 
where $\omega$ is specified in Eq.~\eqref{eq:FIntermediateIntermediate}. (As a technical remark we note that the evaluation of the matching conditions is facilitated by using that after analytic continuation, the matching points have large imaginary parts, $|\mathrm{Im}(\tau_i)|=t/2$, and an approximation $\varphi=\ln(\omega^2\gamma_i^{-1})\pm 2i(\omega\tau-\delta)$ is possible, where the sign is postive/negative for $\tau_{4,3}$/$\tau_{1,2}$.)
The substitution of these configurations into the pre-exponential factor yields 
\begin{align}
    \prod_{i=1}^4 e^{\frac{1}{4}\bar \varphi(\tau_i)}= \frac{4 \omega_0^2\, e^{-\frac{i\pi}{4}-\omega_0 t}}{((1+\alpha_1)(1+\alpha_2)(1+\alpha_1+\alpha_2))^{1/4}},
\end{align}
where Eq.~\eqref{eq:C4Times} was used. Similarly, the substitution of the
stationary solution into the action yields $S[\bar\varphi]=-2M \pi \omega_0^2$. (To
derive this result, the regularization of the solution near the boundary by the
parameter $\lambda$ of Eq.~\eqref{eq:phi_opt} needs to be taken into account. As
before it leads to a cancellation of the formally divergent factor $\gamma t_H$ in
the action.) Finally, we note that the integration of fluctuations around the optimal
configuration leads to a fluctuation determinant independent of the observation
times. This factor cancels against the  identical pre-exponential factor multiplying
the partition sum~\eqref{eq:Z_sem} and may be ignored. Combining terms, we arrive at
Eq.~\eqref{eq:FIntermediateIntermediate}.

\vspace{0.5cm}
\noindent{\bf References}\\
\bibliographystyle{unsrt}

\bibliography{Bibliography}{}





\end{document}